\documentclass[amsmath,amssymb,nofootinbib,notitlepage,superscriptaddress,twocolumn]{revtex4-1}
\pdfoutput=1

\usepackage{aas_macros,mathtools,esint,graphicx,overpic,xcolor,multirow,tabularx}
\usepackage[bottom]{footmisc}
\usepackage[colorlinks=true]{hyperref}
\newcommand{\arccosh}{\mathrm{arccosh}\,}
\newcommand{\arcsinh}{\mathrm{arcsinh}\,}
\newcommand{\V}{\mathrm{V_{PS}}}

\usepackage{xcolor}
\usepackage[normalem]{ulem}

\begin{document}
\preprint{APS/123-QED}
\title{The symplectic geometry of the black hole photon shell}

\author{Judy~Shir}
\email{shir.judy@gmail.com}
\affiliation{
Department of Physics, Faculty of Natural Sciences, University of Haifa, Haifa 3498838, Israel}
\affiliation{Haifa Research Center for Theoretical Physics and Astrophysics, University of Haifa, Haifa 3498838, Israel}
\author{Shahar~Hadar}
\email{shaharhadar@sci.haifa.ac.il}
\affiliation{
Department of Physics, Faculty of Natural Sciences, University of Haifa, Haifa 3498838, Israel}
\affiliation{Haifa Research Center for Theoretical Physics and Astrophysics, University of Haifa, Haifa 3498838, Israel}
\affiliation{
Department of Mathematics and Physics, University of Haifa at Oranim, Kiryat Tivon 3600600, Israel}

\begin{abstract}
The unstably bound, critical null geodesics of the Kerr spacetime form a distinguished class of orbits whose properties govern observables such as the photon ring and the high-frequency component of black-hole ringdown. This set of orbits defines a codimension-two submanifold of the null-geodesic phase space known as the \textit{photon shell}. In this work we investigate the photon shell's intrinsic symplectic geometry. Using the induced symplectic form, we construct the canonical volume form on the shell and compute the differential phase-space volume it encloses as a function of radius -- equivalently, the radial density of states. In the near-extremal limit the photon shell bifurcates into near-horizon and far-region components; we find that approximately $3\%$ of the shell's phase-space volume resides in the near-horizon component.  We also analyze a thickening of the photon shell that includes near-critical orbits, and compute its differential phase-space volume. Beyond their intrinsic theoretical interest, these results may inform the interpretation of high-resolution observations of spinning black holes.
\end{abstract}
\maketitle
\section{Introduction}
\noindent
In the past several years, the Event Horizon Telescope (EHT) collaboration has published the first images of two supermassive black holes (BHs) \cite{event2019first,EHT2022SgrA*1}, that show annuli of light surrounding dark centers. If the emitting region of these BHs is optically thin, general relativity predicts that within the images lies a thin bright ring, the details of which are still unresolved \cite{bardeen1973black,Luminet1979,Virbhadra2000,beckwith2005extreme,johannsen2010testing}. The excess light of this \emph{photon ring} comes from photons that propagate along nearly bound null geodesics, and are deflected by large angles as they orbit at horizon-scale distance from the BH before embarking on their journey to the telescope. The photon ring has an intricate, universal structure determined by the BH's spacetime geometry and is largely independent of astrophysical parameters \cite{gralla2019black, johnson2020universal,hadar2021photon,cederbaum2019geometry}. 
The region where these bound, or critical, null geodesics orbit the BH is called the \emph{photon shell}. Often, this term is used to refer to the spacetime region traced by the critical null geodesics. In this paper, as in \cite{hadar2022holography}, we use the same term to refer to the (closely related) codimension-two submanifold of the null-geodesic phase space describing states (positions and momenta) of photons that are orbiting on a critical light ray. 

The geometric properties of the photon shell of a Kerr BH, considered as a special, critical phase-space submanifold, are the primary focus of this work. 
Using a convenient choice of coordinates on the (four-dimensional) photon shell, we derive the symplectic form induced on it by the canonical symplectic form in the full (six-dimensional) null-geodesic phase space. We then use it to derive the volume form on the photon shell and integrate it along multiple directions. This allows us to uncover novel 
%coordinate-invariant \textcolor{red}{and Hamiltonian preserving} 
properties of the celebrated Kerr solution that are of intrinsic geometric interest. In particular, we compute the differential phase-space volume as a function of the photon-shell radius, canonically quantifying the density of photon states at any radius. We then integrate over radii and compute the total photon shell volume at fixed energy. Our results are invariant under Hamiltonian-preserving coordinate transformations.

As an application of our results, we closely investigate the near-extremal (high-spin) limit. In this special case it is known that the photon shell resides in both the near-horizon \cite{bardeen1999extreme,Bredberg2010} and far regions \cite{gralla2018observational,gates2021photon}. We show analytically and check numerically that, in the extremal limit, the photon shell bifurcates into two branches: one ($\approx97\%$ of the shell) that does not scale to the horizon near extremality, and the second ($\approx3\%$ of the shell) that scales like the innermost (corotating) critical photon orbit. Remarkably, all photon shell radii of intermediate scalings occupy a vanishingly small phase-space volume in the extremal limit.  
Finally, we thicken the photon shell submanifold into a finite-measure slice of the 6D phase space, that represents near-critical null geodesics--those which are relevant for photon ring observations. We define and compute the volume of such near-critical phase-space slices, whose thickness depends on the half-orbit number of the corresponding geodesics.

In addition to its intrinsic geometric interest, our analysis may have several potential applications.
First, it provides a starting point for an approach to time-averaged BH images in which, rather than modeling the source, one models the averaged phase-space photon distribution.
Second, it could offer new insight into features of quasinormal modes (QNMs), such as their asymptotic growth at high frequency \cite{Jaramillo2024} and the spectrum bifurcation near extremality \cite{Yang2013}.
Finally, it may further clarify the structure and role of the conformal symmetries of near-critical high-frequency BH perturbations \cite{hadar2022holography}.

We open in Sec.~\ref{background} by giving a brief review of Kerr null geodesics, and of the photon ring and shell. Next, in Sec.~\ref{The critical phase space volume}, we derive the natural symplectic form on the photon shell and the volume form that descends from it. We subsequently integrate this volume form, computing the (differential) phase-space volumes of different parts of the shell.
In Sec.~\ref{different limits of the critical phase space volume}, we investigate the low- and high-spin limits of our results. Near extremality, we reveal a remarkable bifurcation of the shell into parts with different near-horizon scaling behaviors. Then, in Sec.~\ref{extended phase space volume}, we extend the shell, which describes exactly critical states, by including near-critical photon orbits, defining and computing the volumes of slices of the full phase space that correspond to light rays undergoing a fixed number of half orbits. Finally, in Sec.~\ref{sec:conclusions} we discuss our main results and some potential future applications. Throughout the paper we use relativistic units $G=c=1$.

\section{Preliminaries}\label{background}
\noindent
In this section, we will describe some of the properties of the Kerr spacetime that will be needed for our calculations. We will begin with background on null geodesics in the Kerr geometry. Next, we will focus on a special subset of these null geodesics--the spherical photon orbits which populate the \emph{photon shell}. We will then describe the structure of the \emph{photon ring}--the brightness enhancement expected to be caused by near-photon shell light rays in BH images. 
\subsection{Null geodesics in the Kerr spacetime}
\noindent
In Boyer-Lindquist coordinates $(t,r,\theta,\phi)$ \cite{boyer1967maximal}, the metric of a Kerr BH with mass $M$ and angular momentum $J = aM$, is given by
\begin{align}
    ds^2 &= -\frac{\Delta}{\Sigma}(dt - a \sin^2(\theta) d\phi)^2 + \frac{\Sigma}{\Delta} dr^2 \\&+\Sigma d\theta^2 + \frac{\sin^2(\theta)}{\Sigma} \Big((r^2+a^2)d\phi - a dt \Big)^2~,\nonumber\\
    \Delta(r) &= r^2 - 2 M r+a^2 , \quad \Sigma(r,\theta) = r^2 + a^2 \cos^2\theta~,
\end{align}
and we assume $0 \leq a\leq M$.
The event horizon radius, corresponding to the larger root of $\Delta$, is given by
\begin{align}
    r_h = M + \sqrt{M^2-a^2}~.
\end{align}
The null geodesics of the Kerr geometry can be defined by two constants of motion: the energy-rescaled azimuthal angular momentum $\lambda =\frac{L}{E}$, where $L=p_{\phi}$ and $E=-p_{t}$, and the energy-rescaled Carter constant $\eta =\frac{Q}{E^2}$, where
\begin{align}\label{Carter}
    Q=p_{\theta}^2+(p_{\phi}^2 \csc^2{\theta}-a^2 p_{t}^2)\cos^2{\theta}~.
\end{align}
With these conserved quantities and the null condition $p_{\mu}p^{\mu}=0$ one can obtain four decoupled first order ordinary differential equations for the null geodesics \cite{gralla2020lensing, gralla2020null},
\begin{align}
  \frac{\Sigma}{E} p^{r} &= \pm_r \sqrt{ \mathcal{R}(r)} \,, \\
   \frac{\Sigma}{E} p^{\theta} &= \pm_{\theta} \sqrt{\Theta(\theta)} \,, \\
   \frac{\Sigma}{E} p^{\phi} &= \frac{a}{\Delta}(r^2+a^2-a\lambda)+\frac{\lambda}{\sin^2\theta}-a~,\\
    \frac{\Sigma}{E} p^{t} &= \frac{r^2+a^2}{\Delta}(r^2+a^2-a\lambda) + a(\lambda - a \sin^2\theta)~,
\end{align}
where
\begin{align}\label{radial pot}
    \mathcal{R}(r) &= (r^2+a^2-a \lambda)^2 - \Delta(r) \big(\eta+(\lambda-a)^2 \big)~,\\
    \Theta(\theta) &= \eta + a^2 \cos^2(\theta)-\lambda^2 \cot^2(\theta)~.\label{theta pot}
\end{align}
The functions $\mathcal{R}(r)$ and $\Theta(\theta)$ are the radial and angular potentials. The symbols $\pm_r$, $\pm_\theta$ indicate the sign of $p^r$ and $p^{\theta}$, respectively, and can take the values $\pm1$.
\subsection{The photon shell and ring}
\noindent
In this work we focus on a particular subset of the null geodesics of the Kerr BH: the \emph{critical}, unstably bound photon orbits, which (mathematically) orbit the BH indefinitely at a fixed Boyer-Lindquist radius $\tilde{r}$ \cite{gralla2019black, teo2003spherical}, and do not fall into the BH horizon or escape to infinity; in Sec.~\ref{extended phase space volume} we will expand the discussion to \emph{near-critical} orbits. Here we briefly review the theory of this family of null geodesics.
The condition for the existence of a critical, bound orbit is $\mathcal{R}(\tilde{r}) = \mathcal{R'}(\tilde{r})=0$, which implies the critical conserved quantities $\tilde{\lambda}, \tilde{\eta}$ satisfy the relations
\begin{align} \label{eq:photon shell lambda}
    \tilde{\lambda}(\tilde{r}) &= a + \frac{\tilde{r}}{a} \Big(\tilde{r}- \frac{2 \Delta(\tilde{r})}{\tilde{r}-M} \Big) \, ,\\
    \tilde{\eta}(\tilde{r})  &= \frac{\tilde{r}^3(4a^2M-\tilde{r}(\tilde{r}-3M)^2)}{a^2(\tilde{r}-M)^2}~. \label{eq:photon shell eta}
\end{align}
Solutions exist only in the range of radii
\begin{gather} \label{eq:rtildes}
  \tilde{r}_{-} \leq \tilde{r} \leq \tilde{r}_{+} \, , \\
  \tilde{r}_{\pm} = 2M \left[1+\cos(\frac{2}{3} \arccos(\pm \frac{a}{M})) \right]~,\nonumber
\end{gather}
and satisfy $\mathcal{R''}(\tilde{r})<0$, which implies they are unstably bound: perturbing them slightly yields a trajectory that extends to infinity or the horizon.
On the boundaries $\tilde{r}=\tilde{r}_{\pm}$, the orbits are equatorial and circular, while at intermediate radii their polar angle oscillates between the turning points $\theta_\pm \gtrless \pi/2$,
\begin{gather} 
\theta_{-} \leq   \theta \leq \theta_{+} \,, \nonumber\\
\quad \theta_{\pm} = \arccos({\mp \sqrt{u_+}})~,
\label{eq:thetapm}
\end{gather} 
where
\begin{align}\label{upm}
    u_{\pm} &= \frac{\tilde{r}}{a^2(\tilde{r}-M)^2} \Big( -\tilde{r}^3+3M^2 \tilde{r}-2 a^2 M \\& \pm 2 \sqrt{M \Delta (2\tilde{r}^3-3M\tilde{r}^2+a^2M)} \Big)~. \nonumber
\end{align}
In the Schwarzschild limit $a \to 0$, the spatial locus of bound photon orbits degenerates into a two-dimensional ``photon sphere" at radius $r=3M$.
The region of spacetime defined by \eqref{eq:rtildes}, \eqref{eq:thetapm} and all $(t,\phi)$, was coined the \emph{photon shell} in \cite{johnson2020universal}. This is the region where the bound orbits exist. Here, we will use the same term--photon shell--to refer to the closely related, yet in-principle different concept of the codimension-2 submanifold of phase space corresponding to the critical, bound orbits. 
\begin{figure}[ht]
  \centering
  \begin{minipage}[b]{0.4\textwidth}\label{photon shell}
    \includegraphics[width=\textwidth]{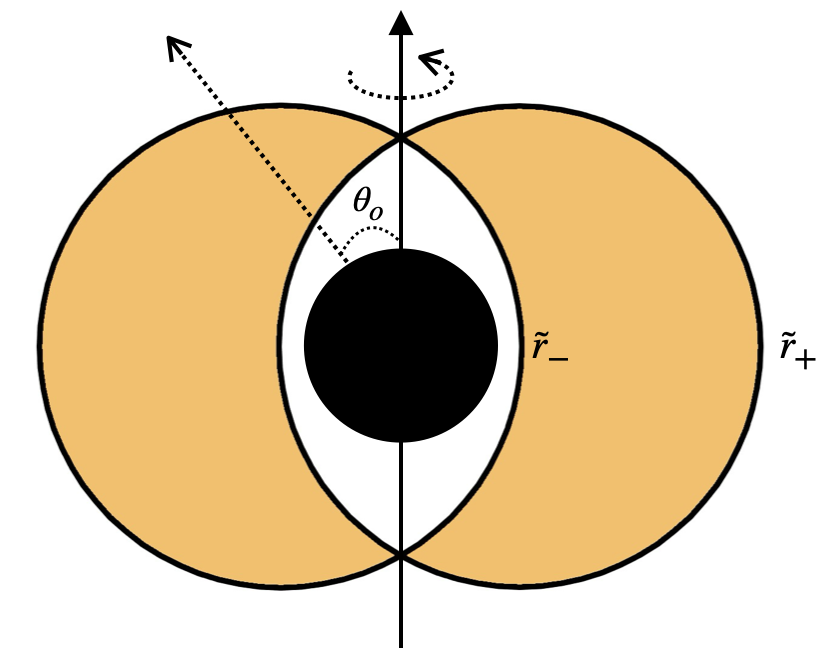}
    \caption{Schematic illustration of the cross section of the Kerr photon shell in the $(r,\theta)$ plane in Boyer-Lindquist coordinates.}
    \label{fig:The photon shell}
  \end{minipage}
    \hfill
  \begin{minipage}[b]{0.4\textwidth}
    \includegraphics[width=\textwidth]{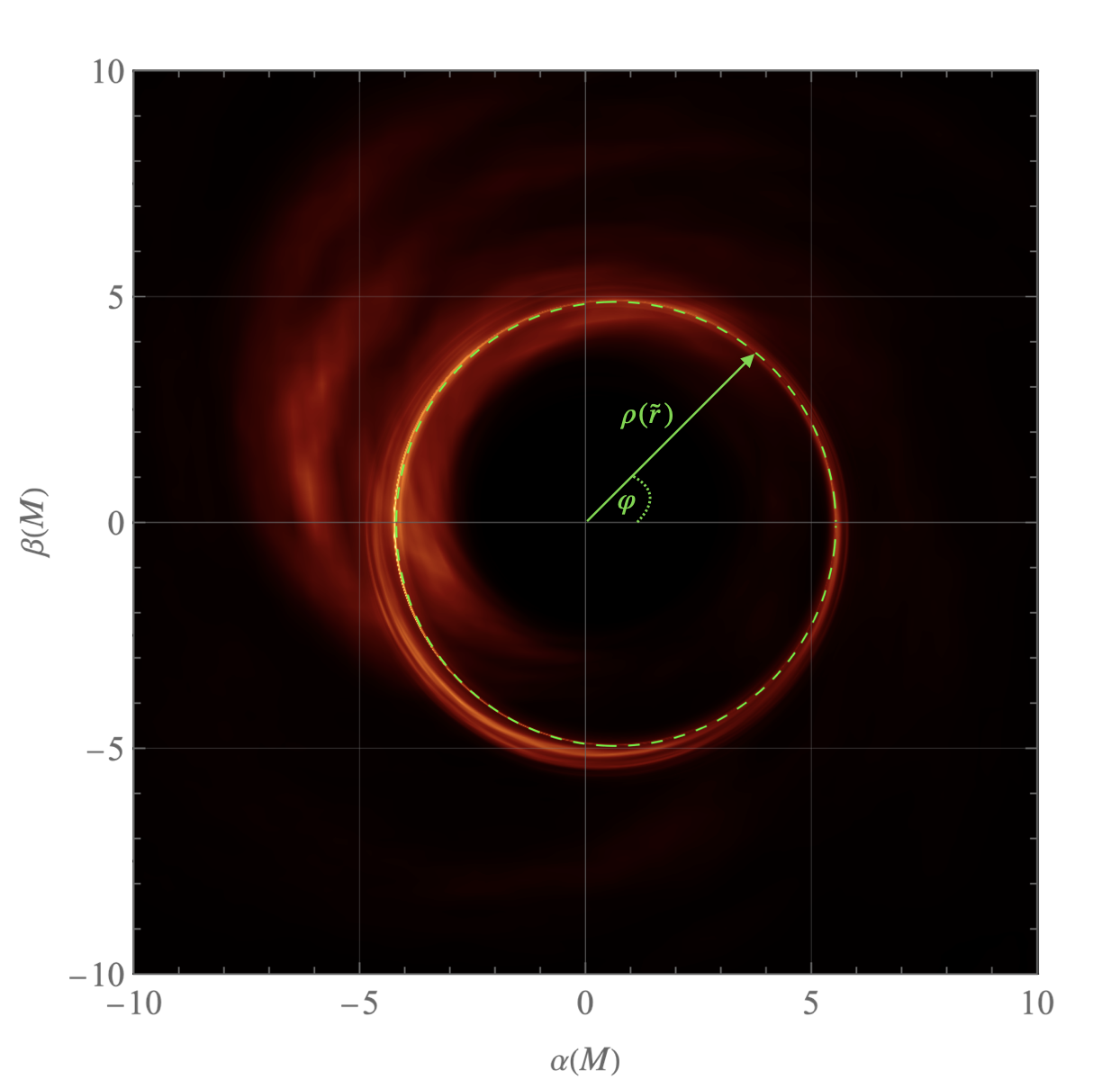}
    \caption{The photon ring is the brightness enhancement in black hole images which follows closely the critical curve $\mathcal{C}$, indicated by the dashed green circle-like line. The background snapshot (excluding the dashed green curve) is taken from the accompanying iAART GitHub repository \cite{iAART}, which is described in \cite{cardenas2023adaptive}.}
    \label{fig:critical curve}
  \end{minipage}
\end{figure}
The critical photon orbits are unstably bound, so when their radial position is slightly perturbed away from $\tilde{r}$ (while keeping the conserved quantities $\tilde{\lambda}, \tilde{\eta}$ fixed) they either fall into the BH or escape to infinity, where they can reach an observer's screen. In that case, their radial deviation from the exactly critical radius, $\delta r=r-\tilde{r}$, grows exponentially with the Lyapunov exponent \cite{johnson2020universal,gralla2020lensing}
\begin{align}\label{eq:Lyapunov}
 \gamma &= \frac{4 \tilde{r} \sqrt{\chi(\tilde{r})}}{a \sqrt{-u_-}} K\left[\frac{u_+}{u_-}\right] \,,
\end{align}
where $\chi(\tilde{r}) = 1-\frac{M \Delta(\tilde{r})}{\tilde{r}(\tilde{r}-M)^2}$ and $K[z]$ is the complete elliptic integral of the first kind.
Such radially perturbed, precisely critical orbits appear on an observer's screen only along a particular closed curve, known as the \emph{critical curve} $\mathcal{C}$, see Fig.~\ref{fig:critical curve}.
Near-critical orbits, which have conserved quantities close to \eqref{eq:photon shell lambda}, \eqref{eq:photon shell eta}, make up the \emph{photon ring} \cite{johannsen2010testing, gralla2020shape, gralla2019black, johnson2020universal, gralla2020measuring}. 
Such orbits can appear on the observer's screen either outside or inside the critical curve $\mathcal{C}$. Tracing the geodesics backward in spacetime from the screen to the far past shows that the ones that appear outside $\mathcal{C}$ asymptote to infinity in the past, and those that appear inside $\mathcal{C}$ asymptote to the horizon in the past. In this way, the critical curve delineates these two distinct families of geodesics.
A light source in an optically thin environment near a BH creates numerous images on the observer's screen. Each image can be labeled by the number of half orbits $n$ executed by the photons on their way from the light source to the telescope. More precisely, $n$ is given by the number of polar turning points the photons composing an image experience\footnote{Interestingly, for sufficiently large spin, inclination, and half-orbit number, there exist multiple images with the same half-orbit number; these images are distinguished by their different azimuthal winding numbers \cite{Zhou2025}.}. 
Generally, the direct ($n=0$) image contributes most of the flux of the total image. Higher-order images are demagnified by $\sim e^{-\gamma}$ per half orbit\footnote{This prediction becomes precise for large $n$ \cite{johnson2020universal}.}, giving rise to a sequence of increasingly thin \textit{subrings} that asymptote to $\mathcal{C}$, each of which is an image of the direct emission. 
These higher-order subrings, which form the photon ring, can also be labeled by a half-orbit number $n\geq1$, and since they are exponentially demagnified, the main contribution to the photon ring flux will come from the $n=1$ subring.
For what follows, it will be useful to recall the relation between the conserved quantities $(\lambda,\eta)$ and coordinates on a far observer's screen \cite{bardeen1973black} at distance $r_o$ and polar inclination $\theta_o$ from the BH spin axis, see Figs.~\ref{fig:The photon shell},\ref{fig:critical curve}. In polar screen coordinates $(\rho, \varphi)$
\begin{align}\label{polar critical curve}
    & \rho = \frac{1}{r_o} \sqrt{a^2(\cos^2{\theta_o}-u_+u_-)+\lambda^2} \,, \\
    & \cos \varphi =-\frac{\lambda}{r_o \rho \sin{\theta_o}}~. \label{varphi critical curve}
\end{align}
Remarkably, the critical curve is naturally parameterized by $\tilde{r}$, defining a map between the radii in the shell and the angles around the ring, $\mathcal{C}=\{\rho(\tilde{r}),\varphi(\tilde{r})\}$ \cite{johnson2020universal}.

\section{The phase-space volume of the photon shell}\label{The critical phase space volume}
\noindent
The state of a particle moving in the Kerr spacetime may be described as a point in the six-dimensional phase space spanned by $(r,\theta,\phi,p_r,p_\theta,p_\phi)$, equipped with the canonical symplectic form
\begin{align}\label{eq:canonical symplectic form}
    \omega = dr \wedge dp_{r} +d\theta \wedge dp_{\theta} +d\phi \wedge dp_{\phi}~.
\end{align}
Time evolution is generated by the Hamiltonian $H=-p_t$, which can be explicitly expressed in terms of the phase-space coordinates by solving $p^\mu p_\mu=0$.
Here, we will be interested in the critical codimension-two submanifold $\mathcal{M}_{\mathrm{PS}}$ of this phase space, which corresponds to exactly bound orbits -- the photon shell. We will take $(\theta,\phi,E,\tilde{r})$ as coordinates on this submanifold, whose embedding in the full phase space is defined by $p_r=0$ and Eqs.~\eqref{eq:photon shell lambda}, \eqref{eq:photon shell eta}. In order to uniquely determine a point in the photon shell, these coordinates must be supplemented by the discrete label $\pm_\theta$, which defines two different sheets that connect along the hypersurfaces $p_\theta=0$.
The induced symplectic form on the photon shell submanifold is the pullback of \eqref{eq:canonical symplectic form} associated with the aforementioned embedding, which yields
\begin{align}\label{eq:critical symplectic form}
   \omega &=  d\theta \wedge \big(\frac{\partial p_{\theta}}{\partial E} dE+ \frac{\partial p_{\theta}}{\partial \tilde{r}} d \tilde{r} \big)\\& \nonumber +d\phi \wedge \big(\frac{\partial p_{\phi}}{\partial E} dE+ \frac{\partial p_{\phi}}{\partial \tilde{r}} d \tilde{r} \big)~,
\end{align}
where
\begin{align}\label{eq:coefficients ptheta}
\frac{\partial p_{\theta}}{\partial E}&=\pm_{\theta} \frac{\sqrt{\bar{\Delta}}}{a (\tilde{r}-M)} \,, \\
\frac{\partial p_{\theta}}{\partial \tilde{r}}&= - \frac{\pm_{\theta}2 E \tilde{r} \chi(\tilde{r})}{a \sqrt{\bar{\Delta}}} \big(a^2 \cot ^2(\theta ) (M+\tilde{r})\\&+\tilde{r}^2 \csc ^2(\theta ) (\tilde{r}-3 M) \big) \,, \\
\frac{\partial p_{\phi}}{\partial E}&=\frac{\tilde{r}^2 (3 M-\tilde{r})-a^2 (M+\tilde{r})}{a (\tilde{r}-M)} \,, \\
\frac{\partial p_{\phi}}{\partial \tilde{r}}&=-\frac{2 E \tilde{r} \chi(\tilde{r}) }{a} \,,\label{eq:coefficients pphi}
\end{align}
for which
\begin{align}
    \bar{\Delta}&=\cos ^2(\theta ) \big(a^4 (\tilde{r}-M)^2-\csc ^2(\theta ) \big(a^2 (M+\tilde{r})
\\& \nonumber +\tilde{r}^2 (\tilde{r}-3 M)\big)^2\big)-\tilde{r}^3 \big(\tilde{r}(\tilde{r}-3M)^2-4 a^2 M \big)~.
\end{align}
Since the critical submanifold is four-dimensional, the volume form on it is given by squaring its symplectic form, 
\begin{align}\label{eq:V PS}
   % dv_{\mathrm{PS}}= \frac{1}{2!} \, \, \omega \wedge \omega  = \Omega_{\mathrm{PS}} \, \,  d\theta \wedge d\phi \wedge dE \wedge  d\tilde{r} ~.\\
   v_{\mathrm{PS}}= \frac{1}{2!} \, \, \omega \wedge \omega  = \Omega_{\mathrm{PS}} \, \,  dE \wedge d\tilde{r} \wedge d\theta \wedge  d\phi ~,
\end{align}
where 
\begin{align} \label{eq:density of states}
   \Omega_{\mathrm{PS}} = \pm_{\theta} \frac{2 E \tilde{r} \chi(\tilde{r})  \left(a^2 \cos ^2(\theta ) (\tilde{r}-M)+\tilde{r}^2 (3 M+\tilde{r})\right)}{\sqrt{\bar{\Delta}}}~,
\end{align}
is a density of states.
We take the unit vectors $\left(\partial/\partial \theta,\partial/\partial \phi,\partial/\partial E,\partial/\partial \tilde{r}\right)$ to be positively oriented on the sheet $\pm_\theta=1$. In order for the manifold to be globally oriented, $\left(\partial/\partial \phi,\partial/\partial \theta,\partial/\partial E,\partial/\partial \tilde{r}\right)$ must be taken to be positively oriented on the sheet $\pm_\theta=-1$, so that volumes on that sheet are also positive. We emphasize that while the overall sign of the shell volume
is a matter of convention, the relative sign between sheets in the definition of their orientation is required to keep the manifold globally oriented, with our coordinate choice.

Eqs.~\eqref{eq:V PS}, \eqref{eq:density of states} provide a prescription for computing volumes in the photon shell which descends from the canonical symplectic form. We will be interested in both the photon shell's overall volume and in what we call \emph{differential phase-space volumes}, which are volumes of regions of infinitesimal extent along one or more of the directions in the critical submanifold. Integrals of differential phase-space volumes yield the total volume, as will be explained below.

\begin{figure}
    \centering
    \includegraphics[width=1\linewidth]{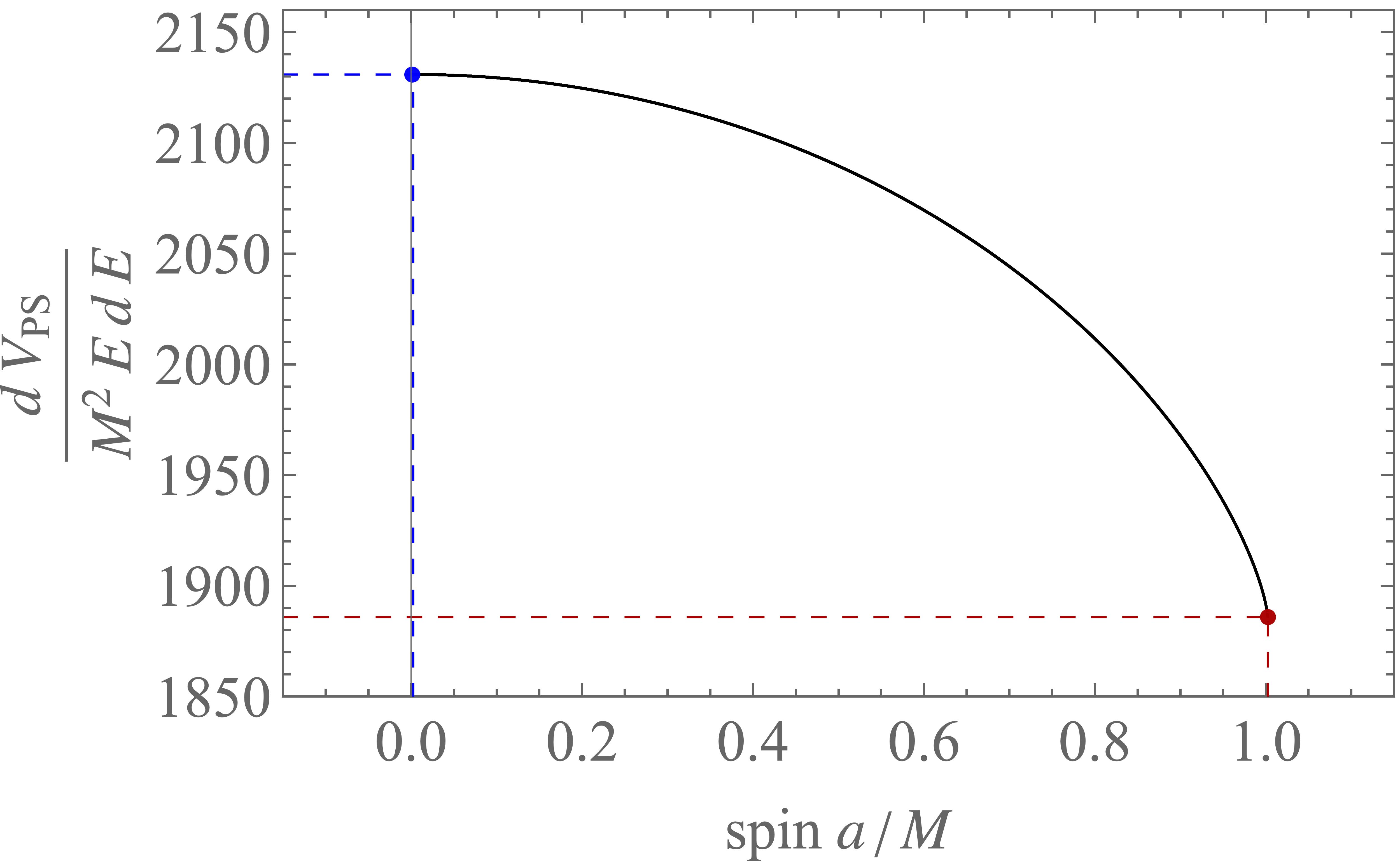}
    \caption{The total (integrated over $\tilde{r}$) differential-in-energy phase-space volume of the photon shell $\frac{dV_\mathrm{PS}}{dE}$ as a function of the spin $a$, at fixed $M$. It is largest at the Schwarzschild limit (blue) and smallest at the near-extremal limit (red).}
    \label{fig:pfinalCritical}
\end{figure}

The total (signed) 4-volume $\mathrm{V}$ of a domain $\mathcal{D}$ in the photon shell is given by 
\begin{align}
%\mathrm{V}=\int_\mathcal{D} dv_\mathrm{PS} \,.\\
\mathrm{V}=\int_\mathcal{D} v_\mathrm{PS} \,.
\end{align}
Clearly, since $\Omega_\mathrm{PS}\sim E$, if $\mathcal{D}$ is taken to be the entire photon shell $\mathcal{M}_{\mathrm{PS}}$, covered by \eqref{eq:rtildes}, \eqref{eq:thetapm}, $0\leq\phi<2\pi$, and $0<E<\infty$ (and $\pm_\theta=\pm 1$), the integral will diverge due to the contribution of arbitrarily high energies -- a ``UV divergence''. In order to make sense of it, one of at least two strategies can be adopted. The first is to introduce an energy cutoff $\Lambda$. This cutoff can be thought of as marking the energy scale where backreaction begins to be significant and perturbation theory breaks down; understanding this regime in detail seems interesting and challenging; it is beyond the scope of the present work. A second possible strategy is to consider the differential (in energy) phase-space volume,
\begin{align}\label{eq:dV/dE}
     %\frac{d\V}{dE} =
     \frac{d\V}{dE} (\bar{E}) &=  \int_{x \in \mathcal{M}_{\mathrm{PS}}, E(x) = \bar{E}} i_{\xi_{E}} (v_\mathrm{PS}) \\ &=  \sum_{\pm_\theta}\int_{\tilde{r}_-}^{\tilde{r}_+}  \int_0^{2\pi} \int_{\theta_-}^{\theta_+} |\Omega_{\mathrm{PS}}(\bar{E})| d\theta d\phi d\tilde{r}  \,. %\\
     %\edit{\left. \frac{d\V}{dE} \right|_{E=\text{const.}} =  \sum_{\pm_\theta}\int_{\tilde{r}_-}^{\tilde{r}_+}  \int_0^{2\pi} \int_{\theta_-}^{\theta_+} i_{\xi_{E}} (v_\mathrm{PS})  \,,}
     % \\
     %\edit{ \frac{d\V}{dE} (\bar{E}) =  \int_{x \in \mathcal{M}_{\mathrm{PS}}, E(x) = \bar{E}} i_{\xi_{E}} (v_\mathrm{PS})  \,,}
\end{align}
%\edit{
where $\xi_{E}$ is a vector field on $\mathcal{M}_{\mathrm{PS}}$, such that the contraction  $i_{\xi_{E}}(dE):=dE(\xi)=1$. This contraction defines a volume form on fixed-$E$ slices of $\mathcal{M}_{\mathrm{PS}}$. In \eqref{eq:dV/dE} we think of $E(x)$ as a function on $\mathcal{M}_{\mathrm{PS}}$, and $\bar{E}$ is the value it is fixed to.
%}
%(or where $\xi_{j}$, $j \in \{E, \tilde{r}\}$ is a vector field on $\mathcal{M}_{\mathrm{PS}}$, such that  $i_{\xi_{j}}(dx^{j}):=dx^j(\xi)=1$.)
Due to the simple energy dependence of \eqref{eq:density of states}, the choice of strategy will only affect subsequent computations quite trivially: the first gives a factor of $\Lambda^2/2$ after integration over energies, while the second yields a factor of $E$. We will be interested also in the differential phase-space volume, in both energy and photon shell radius, defined by
\begin{align}\label{eq:dV/dEdrtilde}
    %\frac{d\V}{dE d\tilde{r}} = 
    \frac{d\V}{dE d\tilde{r}} (\bar{E},\bar{\tilde{r}}) &=  \int_{x \in \mathcal{M}_{\mathrm{PS}}, E(x) = \bar{E}, \tilde{r}(x)=\bar{\tilde{r}}} i_{\xi_{\tilde{r}}} i_{\xi_{E}}  (v_\mathrm{PS}) \\ 
    &= \sum_{\pm_\theta}\int_0^{2\pi} \int_{\theta_-}^{\theta_+} |\Omega_{\mathrm{PS}}(\bar{E},\bar{\tilde{r}})| d\theta d\phi \,,
    %\\
    %\edit{\left. \frac{d\V}{dE d\tilde{r}}\right|_{\substack{E=\text{const.} \\ \tilde{r}=\text{const.}}} = \sum_{\pm_\theta}\int_0^{2\pi} \int_{\theta_-}^{\theta_+} i_{\xi_{\tilde{r}}} i_{\xi_{E}}  (v_\mathrm{PS})\,,}
      %\\
     %\edit{ \frac{d\V}{dE d\tilde{r}} (\bar{E},\bar{\tilde{r}}) =  \int_{x \in \mathcal{M}_{\mathrm{PS}}, E(x) = \bar{E}, \tilde{r}(x)=\bar{\tilde{r}}} i_{\xi_{\tilde{r}}} i_{\xi_{E}}  (v_\mathrm{PS})  \,,}
\end{align}
where the contraction $i_{\xi_{\tilde{r}}}$ is defined analogously to $i_{\xi_{E}}$.
\eqref{eq:dV/dEdrtilde} quantifies the contributions of different photon shell radii to $\frac{d\V}{dE}$. 
%\edit{We can express the differential phase space volumes in terms of the coordinates,
%\begin{align}
%         \frac{d\V}{dE} &=  \sum_{\pm_\theta}\int_{\tilde{r}_-}^{\tilde{r}_+}  \int_0^{2\pi} \int_{\theta_-}^{\theta_+} |\Omega_{\mathrm{PS}}| d\theta d\phi d\tilde{r}  \,,\\
%         \frac{d\V}{dE d\tilde{r}} &= \sum_{\pm_\theta}\int_0^{2\pi} \int_{\theta_-}^{\theta_+} |\Omega_{\mathrm{PS}}(\bar{E},\tilde{r})| d\theta d\phi \,.
%\end{align}}
Evaluating \eqref{eq:dV/dEdrtilde} gives 
%Integrating over $\theta$ and $\phi$ in \eqref{eq:dV/dEdrtilde} and summing over both signs of $p_\theta$ gives 
\begin{align}\label{eq:dVCritical after y int}
\frac{d\V}{dE d\tilde{r}}&= 
\frac{16 \pi  \tilde{r} \chi(\tilde{r})E}{\sqrt{-u_{-}}} \Big[
 u_{-}\left(K \Big[\frac{u_{+}}{u_{-}}\Big]-E\Big[\frac{u_{+}}{u_{-}}\Big]\right) \nonumber \\&+ \frac{\tilde{r}^2 (3 M+\tilde{r})}{a^2 (\tilde{r}-M)}  K\Big[\frac{u_{+}}{u_{-}}\Big] \Big]~,
\end{align}
where $K[z]$ and $E[z]$ are the complete elliptic integral of the first and second kinds, respectively, and we note that $u_{-}<0$.
\begin{figure}[ht]
    \centering
    \includegraphics[width=1\linewidth]{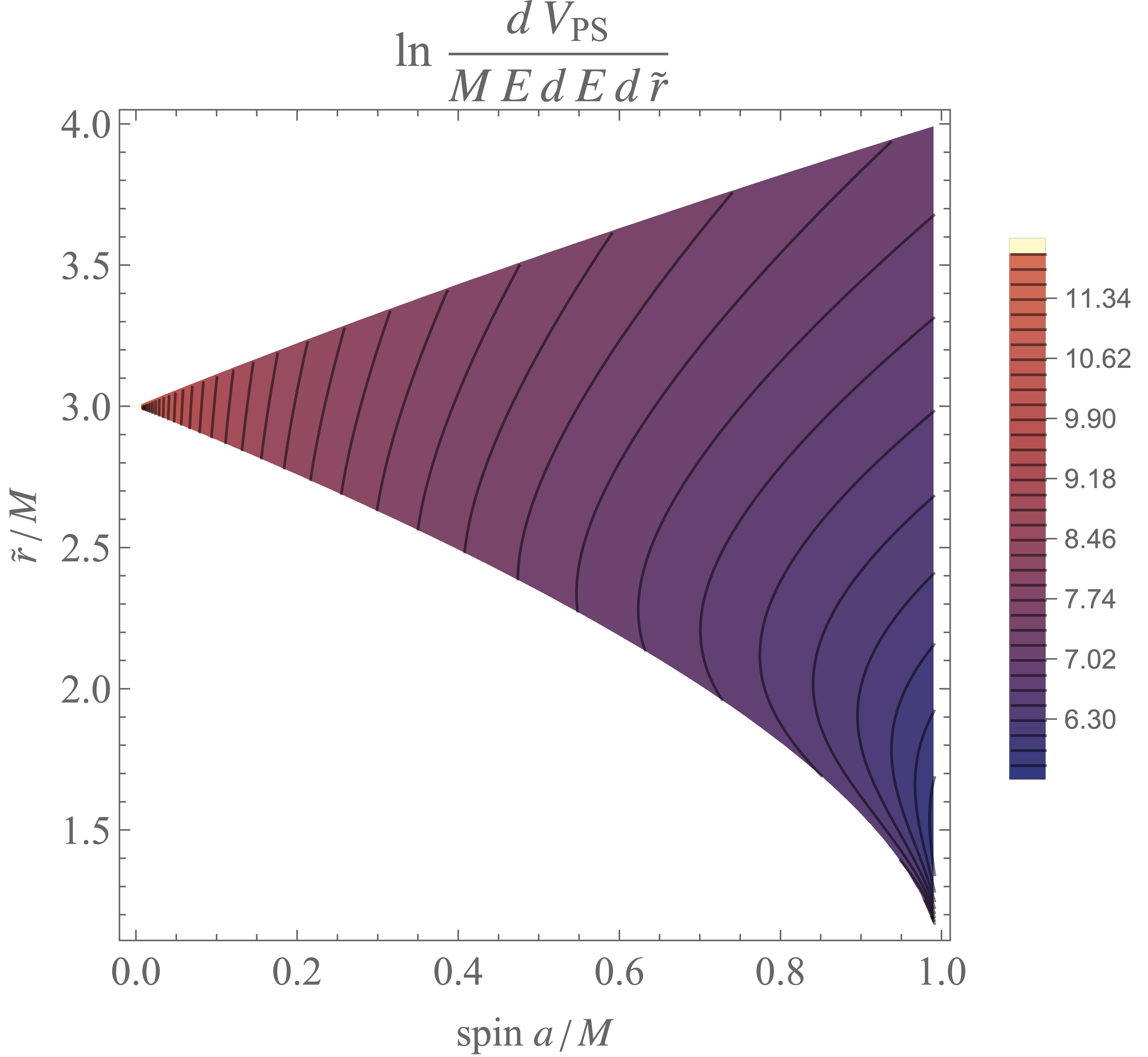}
    \caption{Differential phase-space volume of the Kerr photon shell as a function of the radius $\tilde{r}$ and the spin $a$.
    The differential volume is largest at the Schwarzschild limit where the photon shell is a sphere of radius $\tilde{r}=3M$. The density
depends nontrivially on $\tilde{r}$. As the spin increases, the density reduces and the photon shell region becomes larger. In the extremal limit ($a\to M)$ it extends in the range $M\le \tilde{r} \le 4M$.}
    \label{fig:dVcritical}
\end{figure}
The computation of \eqref{eq:dV/dE} requires another integration, with respect to $\tilde{r}$.
We were unable to find an analytical closed-form expression for the result; its numerical evaluation is nevertheless straightforward.
In Fig.~\ref{fig:dVcritical} we plot the differential volume of the critical submanifold as a function of the spin $a$ and the photon shell radius $\tilde{r}$.
In Fig. \ref{fig:pfinalCritical} we plot the total (integrated over $\tilde{r}$) critical phase-space volume as a function of the spin $a$, with fixed BH mass $M$. The critical phase-space volume is largest at the Schwarzschild limit ($a=0$) and then decreases as it approaches the near-extremal limit ($a \to M$). 
In  Sec.~\ref{different limits of the critical phase space volume}, 
we will study in more detail the low- and high-spin limits of the differential phase-space volume of the photon shell. 

\subsection{Invariance of the differential phase space volume under Hamiltonian-preserving changes of time foliation}\label{Invariance}
In this section we demonstrate that the differential volumes $\frac{d\V}{dE}$, $\frac{d\V}{dE d\tilde{r}}$, defined in Eqs.~\eqref{eq:dV/dE},~\eqref{eq:dV/dEdrtilde}, are invariant under changes of time foliation, i.e., spacetime coordinate transformations of the form
\begin{align} \label{eq: time foliation change}
    \bar{t} &= t - f(r,\theta, \phi) \quad ; \quad x^{\bar{i}} = x^i~,
\end{align}
where $ i \in \{r,\theta,\phi\}, \bar{i} \in \{\bar{r},\bar{\theta},\bar{\phi}\}$ and $f(r,\theta, \phi)$ is a differentiable function of the spatial coordinates. We consider only time-independent foliations, that explicate Kerr's Killing symmetry of stationarity, keeping the metric and Hamiltonian naturally time-independent.
\eqref{eq: time foliation change} induce the momenta transformations
\begin{align}\label{eq: momentum transformation}
    p_{\bar{t}}=p_t \quad ; \quad p_{\bar{i}}=p_i+p_t \partial_i f~.
\end{align}
Computing the canonical symplectic 2-form in the new coordinate system gives
\begin{align}\label{eq: new symplectic 2-form}
    \bar{\omega}=\omega +df \wedge dp_t~,
\end{align}
where $\omega$ is given in Eq.~\eqref{eq:canonical symplectic form} and $p_t=-E$. The additional term is given by
\begin{align}
    df \wedge dp_t &=-( \frac{\partial f}{\partial \tilde{r}}d \tilde{r}\wedge dE+\frac{\partial f}{\partial \theta} d\theta \wedge dE +\frac{\partial f}{\partial \phi} d\phi \wedge dE)~.
\end{align}
Squaring the new symplectic form gives
\begin{align}\label{eq:new volume form}
    \bar{\omega} \wedge \bar{\omega} = \omega \wedge \omega - 2\omega \wedge df \wedge dE~,
\end{align}
where,
\begin{align}
   \omega \wedge df \wedge dE &= \Big( \frac{\partial p_{\theta}}{\partial \tilde{r}} \frac{\partial f}{\partial \phi}  -\frac{\partial p_{\phi}}{\partial \tilde{r}} \frac{\partial f}{\partial \theta}\Big) dE \wedge d\tilde{r} \wedge d\theta \wedge d \phi~. 
\end{align}
Contracting Eq.~\eqref{eq:new volume form} with $i_{\xi_{E}},i_{\xi_{\tilde{r}}}$ gives
\begin{align} \label{eq: barred differential PSV}
&\frac{d\mathrm{\bar{V}}_\mathrm{PS}}{dE d\tilde{r}}(\bar{E},\bar{\tilde{r}}) = \frac{d\V}{dE d\tilde{r}}(\bar{E},\bar{\tilde{r}}) \\ 
&-2\int_{x \in \mathcal{M}_{\mathrm{PS}} , E(x) = \bar{E}, \tilde{r}(x)=\bar{\tilde{r}}} \Big(\frac{\partial p_{\theta}}{\partial \tilde{r}} \frac{\partial f}{\partial \phi} -\frac{\partial p_{\phi}}{\partial \tilde{r}} \frac{\partial f}{\partial \theta}  \Big)d\theta \wedge d \phi  \,, 
%    \frac{d\mathrm{\bar{V}}_\mathrm{PS}}{dE d\tilde{r}} &= \frac{d\V}{dE d\tilde{r}} \\ 
%    &-2 \sum_{\pm_\theta}\int_0^{2\pi} \int_{\theta_-}^{\theta_+}\Big(\frac{\partial p_{\theta}}{\partial \tilde{r}} \frac{\partial f}{\partial \phi} -\frac{\partial p_{\phi}}{\partial \tilde{r}} \frac{\partial f}{\partial \theta}  \Big)d\theta \wedge d \phi 
\end{align}
where $\frac{d\V}{dE d\tilde{r}}$ is given in Eq.~\eqref{eq:dV/dEdrtilde}.
The domain of integration is the codimension-2 (equivalently, 2D) submanifold of $\mathcal{M}_{\mathrm{PS}}$ defined by fixing specific values for the conserved momenta $E=\bar{E}$ and $\tilde{r}=\bar{\tilde{r}}$. 
By the Liouville-Arnold theorem \cite{arnold1989mathematical} (for a discussion in the Kerr context, see \cite{schmidt2002celestial}), this domain is diffeomorphic to the 2-dimensional torus $\mathcal{T}^2$, and is in particular a closed manifold.
In terms of coordinates, this manifold is spanned by
\begin{align}
\{ (\theta,\phi,\pm_\theta) \mid \theta_- \le \theta \le \theta_+, \; 0 \le \phi \le 2\pi, \; \pm_\theta \in \{\pm 1\} \},
\end{align}
together with the identifications $\phi \sim \phi +2 \pi$ and $\theta_\pm|_{\pm_\theta=+} \sim \theta_\pm|_{\pm_\theta=-}$. 
The integrand in the second term on the RHS in \eqref{eq: barred differential PSV} can be written in the following way:
\begin{align}
   \Big(\frac{\partial p_{\theta}}{\partial \tilde{r}} \frac{\partial f}{\partial \phi} &-\frac{\partial p_{\phi}}{\partial \tilde{r}} \frac{\partial f}{\partial \theta}  \Big) d\theta \wedge d \phi  %\\ 
   %&= -d[\frac{\partial p_{\theta}(\theta)}{\partial \tilde{r}} f(\phi)d\theta] -\frac{\partial p_{\phi}}{\partial \tilde{r}} d[f(\theta)d\phi] \,.
    \\ &= -d[\frac{\partial p_{\theta}}{\partial \tilde{r}} f d\theta + \frac{\partial p_{\phi}}{\partial \tilde{r}} f d\phi] \,.
\end{align}
We assume that $f$ is a smooth, globally single-valued function on $\mathcal{T}^2$. We therefore need to verify that the full one-form inside the exterior derivative is globally well-defined.
The second term is smooth and globally single-valued, since $\partial p_{\phi}/\partial\tilde r$, given in Eq.~\eqref{eq:coefficients pphi}, is well-defined and single-valued. Although the angle $\phi$ is not globally single-valued, $\phi\sim\phi+2\pi$, its differential $d\phi$ is globally defined. Hence,
\begin{align}
    \frac{\partial p_{\phi}}{\partial\tilde r}f\,d\phi
\end{align}
is a globally defined one-form on $\mathcal{T}^2$.
We next consider the first term. Since
\begin{align}
    p_\theta=\pm_\theta E\sqrt{\Theta(\theta,\tilde r)},
\end{align}
we have
\begin{align}
\label{eq: derivative of p_theta by rt}
    \frac{\partial p_{\theta}}{\partial \tilde{r}}
    =
    \frac{\pm_{\theta}E}{2\sqrt{\Theta(\theta,\tilde r)}}
    \frac{\partial\Theta(\theta,\tilde r)}{\partial\tilde r}.
\end{align}
Thus, $\partial p_\theta/\partial\tilde r$ is singular at the polar turning points $\theta_\pm$, where $\Theta=0$. However, this singularity is a coordinate singularity and it can be removed by introducing the polar orbit coordinate $n$ \cite{gralla2020lensing}. Note that $n$ here is a continuous coordinate defined by
\begin{align}
    dn
    =
    \frac{\pm_\theta\,d\theta}
    {G_\theta^1\sqrt{\Theta(\theta,\tilde r)}},
\end{align}
where $G_\theta^1$ is a constant defined in Eq.~\eqref{eq: G_theta one half orbit}
, and the integral part of $n$ is, by definition, the half-orbit number mentioned above. 
%Using Eqs.~\eqref{eq: G_theta}, \eqref{eq: definition of n}, and \eqref{eq: G_theta one half orbit}, we have
Consequently,
\begin{align}
    \frac{\partial p_{\theta}}{\partial\tilde r}f\,d\theta
    &=
    \frac{G_\theta^1 E f}{2}
    \frac{\partial\Theta(\theta(n),\tilde r)}{\partial\tilde r}\,dn.
\end{align}
The factors $\sqrt{\Theta}$ and $\pm_\theta$ have canceled, showing explicitly that the apparent singularity at the turning points is absent from the one-form $(\partial p_\theta/\partial\tilde r)d\theta$.
The quantity $G_\theta^1$ depends only on $\tilde{r}$
%the parameters of the photon orbit 
and is finite. It remains to check the regularity of $\partial\Theta/\partial\tilde r$. Using
\begin{align}
    \frac{\partial\Theta(\theta(n),\tilde r)}{\partial\tilde r}
    =
    \frac{\partial\tilde\eta}{\partial\tilde r}
    -
    2\tilde\lambda
    \frac{\partial\tilde\lambda}{\partial\tilde r}
    \cot^2\theta(n),
\end{align}
and noting that at the turning points
\begin{align}
    \cos^2\theta_\pm=u_+,
\end{align}
we obtain
\begin{align}
    \cot^2\theta_\pm=\frac{u_+}{1-u_+},
\end{align}
which is finite where $0<u_+<1$. Hence, $\partial\Theta(\theta(n),\tilde r)/\partial\tilde r$
%\begin{align}
%    \frac{\partial\Theta(\theta(n),\tilde r)}{\partial\tilde r}
%\end{align}
remains finite at the turning points.
The coordinate $n$ is periodic, $n\sim n+1$, and is therefore an angular coordinate that, together with $\phi$, parametrizes the invariant torus $\mathcal{T}^2$.
%on the polar orbit. 
Although $n$ itself is not single valued, its differential $dn$ is globally defined, in direct analogy with the azimuthal coordinate $\phi$ and its globally defined differential $d\phi$. Moreover, $\theta(n)$ is periodic under $n\to n+1$. It follows that
\begin{align}
    \frac{G_\theta^1 E f}{2}
\frac{\partial\Theta(\theta(n),\tilde r)}{\partial\tilde r}\,dn
\end{align}
is a smooth, globally defined one-form on the polar orbit.
Thus, the complete one-form
\begin{align}
\frac{\partial p_{\theta}}{\partial\tilde r} f\,d\theta
+
\frac{\partial p_{\phi}}{\partial\tilde r} f\,d\phi
\end{align}
is globally defined and smooth on $\mathcal{T}^2$. Consequently, the two-form is globally exact,
\begin{align}
    \Bigg(\frac{\partial p_{\theta}}{\partial \tilde{r}}
\frac{\partial f}{\partial \phi}
-
\frac{\partial p_{\phi}}{\partial \tilde{r}}
\frac{\partial f}{\partial \theta}
\Bigg)d\theta\wedge d\phi = -d[\frac{\partial p_{\theta}}{\partial \tilde{r}} f d\theta + \frac{\partial p_{\phi}}{\partial \tilde{r}} f d\phi] \,.
\end{align}
Therefore, by Stokes’ theorem, the second term on the RHS in Eq.~\eqref{eq: barred differential PSV} vanishes, demonstrating the invariance of the differential phase-space volume, $\frac{d\mathrm{\bar{V}}_\mathrm{PS}}{dE d\tilde{r}} = \frac{d\V}{dE d\tilde{r}}$, under transformations of the form \eqref{eq: time foliation change}.
In a similar manner, contracting Eq.~\eqref{eq:new volume form} with $i_{\xi_E}$ yields 
%Eq.~\eqref{eq:dV/dE}, together with the additional term, \textcolor{blue}{SH: maybe modify the equation below similarly to those above?}
\begin{align}
  &\frac{d\mathrm{\bar{V}}_\mathrm{PS}}{dE}(\bar{E}) = \frac{d\V}{dE}(\bar{E}) \\ 
&-2\int_{x \in \mathcal{M}_{\mathrm{PS}} , E(x) = \bar{E}} \Big(\frac{\partial p_{\theta}}{\partial \tilde{r}} \frac{\partial f}{\partial \phi} -\frac{\partial p_{\phi}}{\partial \tilde{r}} \frac{\partial f}{\partial \theta}  \Big)d\tilde{r} \wedge d\theta \wedge d \phi  \,, 
  %&\frac{d\mathrm{\bar{V}}_\mathrm{PS}}{dE } = \frac{d\V}{dE }\\&-2 \sum_{\pm_\theta}\int_{\tilde{r}_-}^{\tilde{r}_+} \int_0^{2\pi} \int_{\theta_-}^{\theta_+}\Big(\frac{\partial p_{\theta}}{\partial \tilde{r}} \frac{\partial f}{\partial \phi} -\frac{\partial p_{\phi}}{\partial \tilde{r}} \frac{\partial f}{\partial \theta}  \Big)d\tilde{r} \wedge d\theta \wedge d \phi ~,
\end{align}
where we can express the integrand in the second term on the RHS as
\begin{align}
   - d [\frac{\partial p_{\theta}}{\partial \tilde{r}} f d\theta \wedge d\tilde{r} +\frac{\partial p_{\phi}}{\partial \tilde{r}} f d\phi\wedge d\tilde{r}]~.
\end{align}
As shown in \cite{cederbaum2019geometry}, fixed-$E$ slices of $\mathcal{M}_{\mathrm{PS}}$ have topology $SO(3)$, and are therefore closed.
Invoking Stokes' theorem again, the $f$-dependent contribution vanishes.
We thus conclude that the differential phase space volumes in Eqs.~\eqref{eq:dV/dE} and~\eqref{eq:dV/dEdrtilde} are invariant under Hamiltonian-preserving changes of time foliation.
\section{Low- and high-spin limits of the critical phase-space volume}\label{different limits of the critical phase space volume}
\subsection{The Schwarzschild limit} \label{sec: schw limit of critical}
\noindent
The Schwarzschild limit of the photon shell phase-space volume must be taken with care.
One cannot simply plug in $a=0$ in \eqref{eq:dV/dE} since the $\tilde{r}$ integral degenerates in this limit; rather, it is necessary to first consider a slowly rotating BH, and then take the $a\to 0$ limit. This can be done by expanding the integrand in \eqref{eq:dV/dE}, as well as the integration limits, to linear order in spin. Since in the small-$a$ limit $\tilde{r}_\pm \approx 3M \pm 2a/\sqrt{3}$ to linear order, one may write $\tilde{r} = 3M+\Delta \tilde{r}$, where $\Delta \tilde{r} = \mathcal{O}(a)$ for all radii in the photon shell.
Carrying out this expansion, integrating, and taking $a \to 0$ yields
\begin{align}
   \left.\frac{dV^{\mathrm{Schw}}_\mathrm{PS}}{dE}\right|_{a=0}= 216 \pi^2 M^2 E~.
\end{align}
This expression can be interpreted as follows. A factor of $4\pi (3M)^2$ may be attributed to the spacetime area of the photon sphere at $r=3M$, whereas a factor of $2\pi$ corresponds to integration over the angle parameterizing the allowed directions of photon momentum within the photon sphere. The remaining factor of $3E$ can be associated with a magnitude of the photon's momentum. This momentum must scale $\sim E$ on general grounds, but we are not aware of an independent (of our computation) way to deduce the numerical factor $3$.
\subsection{The (near-)Extremal limit} \label{sec:near-extremal}
\noindent
The critical phase space shows interesting behavior in the near-extremal limit, where the BH spin parameter approaches its maximal allowed value $a=M$. We parameterize the closeness to extremality by $\kappa \equiv \sqrt{1-(a/M)^2} >0$, with $\kappa \ll 1$ near extremality. 
It is well known that the near-horizon geometry of the extremal Kerr spacetime is nondegenerate. This geometry is not resolved by Boyer-Lindquist coordinates, but it can be by introducing coordinates that scale to the horizon, yielding the near-horizon extreme Kerr (NHEK) metric \cite{bardeen1999extreme}.
A related scaling limit, taken by scaling $r\to r_h$ at the same rate as taking $\kappa \to 0$, yields the near-NHEK geometry \cite{Bredberg2010,hadar2014gravity}.
Similar scaling transformations have proved useful in the context of the optical image of near-extreme Kerr null geodesics; namely, in order to understand the structure of the critical curve in near-extreme Kerr \cite{gralla2018observational,gates2021photon}. Here, we will employ a similar strategy in order to resolve the photon shell of near-extreme Kerr.
For $\kappa \ll 1$ the Boyer-Lindquist radii of the outermost/innermost photon shell orbits satisfy $\tilde{r}_+ = 4M +\mathcal{O}(\kappa)$ and $\tilde{r}_- = M+\frac{2}{\sqrt{3}}\kappa+\mathcal{O}(\kappa^2)$, and in particular $\tilde{r}_-$ tends to the Boyer-Lindquist coordinate value of the event horizon. It is understood then that the near-extreme Kerr photon shell occupies both near-horizon and far-region radii. But, precisely, how is the photon shell distributed between the different regions?
Interestingly, as we show below, the (near-)extremal Kerr photon shell bifurcates toward extremality. Approximately $3\%$ of it resides within radii $\tilde{r}-r_h \sim \mathcal{O}(\kappa^1)$ while the remaining $97\%$ reside in the far region, $\tilde{r}-r_h \sim \mathcal{O}(\kappa^0)$; intermediate radial scalings contribute a vanishingly small differential phase-space volume to the photon shell in the limit $\kappa \to 0$. In Fig.~\ref{fig:pBandsPlotCritical1} we illustrate this behavior.
\subsubsection{The near-NHEK region:}
\noindent
In the near-NHEK limit, we take the BH to be near-extremal while zooming in on its near-horizon region at the same rate as extremality is approached. This is achieved by transforming
\begin{align}\label{eq:rt near NHEK transformation}
    \tilde{r} &\to  M(1+\kappa \tilde{R})~,
\end{align}
where $\tilde{R}$ is a dimensionless near-horizon radius in the range $\tilde{R} \in \left[\tilde{R}_{-}=2/\sqrt{3},\infty \right)$, and taking $\kappa \ll 1$.
Plugging into Eq.~\eqref{eq:dVCritical after y int}, to leading order in $\kappa$ we obtain
\begin{align}
    &\frac{dV_\text{near-NHEK}}{dE d\tilde{R}} = \frac{ 64 \pi E M^2 K\big[\frac{Y_+}{Y_-} \big]}{\tilde{R}^3 \sqrt{-Y_-}}~,
\end{align}
where
\begin{align}
    Y_{\pm} = -3+2\tilde{R}^{-2}\pm 2 \sqrt{3-4\tilde{R}^{-2}+\tilde{R}^{-4}} \,.
\end{align}
Integrating over near-NHEK radii, the near-horizon contribution to the photon shell phase-space volume is
\begin{align}
       \left. \frac{dV}{dE}\right|_{\text{near-NHEK}} = \int_{2/\sqrt{3}}^{\infty} \frac{dV_{\text{near-NHEK}}}{dE d\tilde{R}} d\tilde{R}~.
\end{align}
\subsubsection{The far region}
\noindent
In the far region limit, we simply set $\kappa = 0$ (or $a=M$) in Eq.~\eqref{eq:dVCritical after y int} and obtain
\begin{align}
&\frac{dV_{\mathrm{far}}}{dE d\tilde{r}}=\frac{16 \pi  E}{\sqrt{-\Upsilon_-}} \left\{\frac{\tilde{r}^2(3M+\tilde{r})}{M^2}K\Big[\frac{\Upsilon_+}{\Upsilon_-}\Big] \right. \nonumber \\
& \left. + \Upsilon_{-(a=M)}(\tilde{r}-M)\left[K\Big[\frac{\Upsilon_+}{\Upsilon_-}\Big]-E\Big[\frac{\Upsilon_+}{\Upsilon_-}\Big] \right] \right\} \,, \nonumber \\
\end{align}
where $\Upsilon_\pm=u_{\pm}(a=M)$ and we have $\Upsilon_-<0$. Using the Boyer-Lindquist extremal values of the outermost/innermost photon shell radii, $\tilde{r}_+|_{a=M}=4M$, $\tilde{r}_-|_{a=M}=M$, the far-region contribution to the photon shell phase-space volume is given by
\begin{align}
       \left. \frac{dV}{dE}\right|_{\text{far}} = \int_{M}^{4M} \frac{dV_{\mathrm{far}}}{dE d\tilde{r}} d\tilde{r}~.
\end{align}
\subsubsection{Bifurcation of the extreme Kerr photon shell} 
\begin{figure}[ht]
    \centering
    \includegraphics[width=1\linewidth]{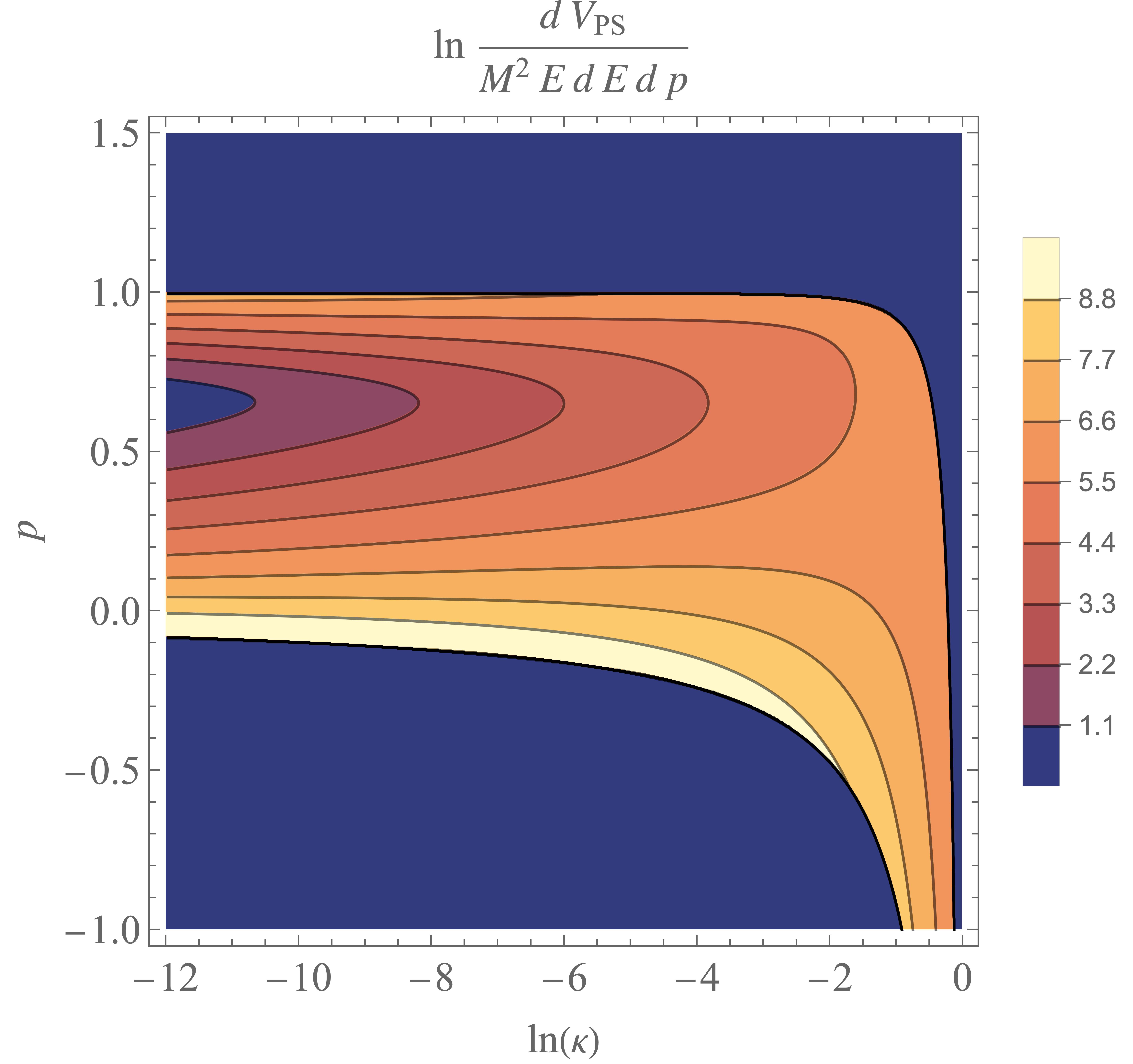}
    \caption{
    Differential phase-space volume of the near-extremal Kerr photon shell as a function of the proximity to extremality and to the horizon. The horizontal axis parameterizes the deviation from extremality via $\ln \kappa$, where $\kappa=\sqrt{1-a^2/M^2}$. The vertical axis parameterizes the distance from the horizon via a ``NHEK-band'' radial coordinate, which we choose to define here as $\tilde{r}=M(1+\frac{2}{\sqrt{3}}\kappa^p)$ so that $p=1$ corresponds to $\tilde{r}_- + \mathcal{O}(\kappa^2)$. The colorscale quantifies the logarithm of the differential (in $p$ and $E$) phase-space volume $\frac{d\V}{dE dp}$. As extremality is approached, moving left on the diagram, the phase
space bifurcates into two branches: the near-NHEK region ($p=1$) and the far region ($p=0$), and a gap opens up between these branches.}
    \label{fig:pBandsPlotCritical1}
\end{figure}

\noindent So far we have analyzed the differential volumes of the near-NHEK and far regions, whose photon-shell radii scale as $\kappa^1$ and $\kappa^0$, respectively. Since the photon shell continuously connects these two regions for any $\kappa>0$, it is natural to ask to what extent intermediate radii contribute to its phase-space volume.
The Boyer-Lindquist radii of spherical (fixed-$r$) orbits in the near-extremal Kerr spacetime can be labeled by
how their radial separation from the horizon $r-r_h$ scales with $\kappa$, the closeness to extremality\footnote{More precisely, the reference here is to a one-parameter family of orbits, each residing in a near-extreme Kerr spacetime with different $\kappa$.}. 
In this context, scaling laws of the type $\tilde{r}-r_h \sim \kappa^p$, with $0<p<1$, often emerge for special orbits. The exponent $p$ is sometimes referred to as the \textit{NHEK band} \cite{gates2021photon, kapec2019particle, gralla2018observational,hadar2014gravity}. For example, as discussed above,
the far region has $p=0$, while the near-NHEK region has $p=1$; another example is the innermost stable circular orbit of massive particles, which has $p=2/3$. 
In the present context, we find that in the extremal limit the contributions of the NHEK bands with $0<p<1$ are vanishingly small. In App.~\ref{app: analytical bifurcation} we show analytically that if near extremality $\tilde{r}_1-r_h \sim \kappa^p$ and $\tilde{r}_2-r_h \sim \kappa^q$, with $0<q<p<1$, then
\begin{align}
    \int_{\tilde{r}_1}^{\tilde{r}_2} \frac{d\V}{dE d\tilde{r}} d\tilde{r} ~~ \xrightarrow[\kappa \to 0] ~~ 0 \,.
\end{align}
We interpret this fact as a bifurcation of the extreme Kerr photon shell. Only the $p=1$ (near-NHEK) and $p=0$ (far) bands contribute to the phase-space volume in the near-extremal limit, and a gap opens up in between. Thus, in the near-extremal limit,
\begin{align} \label{eq:total extremal volume}
 \left.\frac{dV}{dE}\right|_{\text{Kerr}} = \left.\frac{dV}{dE}\right|_{\text{near-NHEK}} + \left.\frac{dV}{dE}\right|_{\text{far}} ~. \nonumber \\
\end{align}
This situation can also be verified by integrating Eq.~\eqref{eq:dVCritical after y int} numerically, for $\kappa\ll1$, and comparing with the near-extremal analytical results. Our analytical and numerical calculations agree that the near-NHEK region constitutes about 2.88\% of the total critical phase-space volume and the far region constitutes about 97.12\% of the volume, such that the two parts sum up to 100\%. In Fig.~\ref{fig:pBandsPlotCritical1} we show the bifurcation of the Kerr photon shell into far and near-NHEK parts, with a gap opening between them as extremality is approached. In this plot, we translate between the Boyer-Lindquist radius and the NHEK band $p$ via the relation $\tilde{r}=M(1+\frac{2}{\sqrt{3}}\kappa^p)$, so that the $p=1$ horizontal line approximates $\tilde{r}_-$ up to $\mathcal{O}(\kappa^2)$ corrections. In the figure, we display the quantity $\frac{d\V}{dE d\tilde{r}} \cdot \frac{d\tilde{r}}{dp}$, where the latter factor is the Jacobian associated with the transformation from $\tilde{r}$ to $p$. In this way, the magnitude of the displayed function directly conveys the total phase-space volume it contributes upon integration over $p$.

\section{The near-critical phase-space volume}\label{extended phase space volume}
\noindent
In the previous sections, we studied the phase-space volume for the null geodesics which are exactly critical. These are photon orbits which, mathematically, orbit the BH forever and neither fall across the event horizon nor escape to infinity. The relevant geodesics for photon ring observations, however, are those which can execute multiple half-orbits around the BH before escaping to infinity. 
Such near-critical null geodesics can be described as slightly displaced in phase space from the submanifold of exactly critical null geodesics--the photon shell. Therefore, in order to study the near-critical phase space, we will add two extra dimensions which thicken the 4D photon shell to a slice of the 6D phase space which is of small but finite thickness. In fact, such a thickening is defined by our choice of a half-orbit number $n$ labeling the slice, and as we will see below, the volume of such a slice scales $\sim e^{-2\gamma n}$.
\subsection{The near-critical phase-space volume element}
\noindent
We find it convenient to parameterize the near-critical phase space by working with the screen polar coordinates $(\rho, \varphi)$, Eqs.~\eqref{polar critical curve}, \eqref{varphi critical curve}, setting $r_o =1$ and $\theta_o=\frac{\pi}{2}$, such that $\rho$ has units of length. We emphasize that we use this relation just as a convenient coordinate choice in phase space, even though we do not discuss optical images here. We parameterize small deviations from the locus of critical conserved quantities, defined by \eqref{eq:photon shell lambda}, \eqref{eq:photon shell eta}, by introducing a shift in $\rho$ while keeping $\varphi$ fixed,
\begin{align}\label{eq: delta rho def}
    \rho \to \tilde{\rho}(\tilde{r})+\delta\rho; \quad     \delta\rho \ll M~,
\end{align}
where $\tilde{\rho}(\tilde{r})$ depends on the conserved quantities $\tilde{\lambda},\tilde{\eta}$ which are functions of the photon shell radius $\tilde{r}$. 
In addition, we define a radial deviation $\delta r$:
\begin{align}\label{radial deviation}
    r \to \tilde{r}+ \delta r; \quad \delta r \ll M~.
\end{align} 
\eqref{eq: delta rho def} allows us to express conserved quantities $\lambda,\eta$, which are close to $\tilde{\lambda},\tilde{\eta}$, using $\delta \rho$:
\begin{align}\label{eta deviation}
    \eta \to \tilde{\eta}(\tilde{r})+ \eta_1(\tilde{r}) \delta\rho +O(\delta \rho^2) \,, \\
   \lambda \to \tilde{\lambda}(\tilde{r})+ \lambda_1(\tilde{r}) \delta\rho +O(\delta \rho^2) \,, \label{lambda deviation}
\end{align}
where:
\begin{align}
    \lambda_1 &= -\frac{2 f(\tilde{r}) \chi(\tilde{r})}{\Delta(\tilde{r})} \Big(\frac{\tilde{r}^2(\tilde{r}-3M)+a^2(\tilde{r}+M)}{a(\tilde{r}+3M)} \Big) \,, \\
    \eta_1 &=\frac{4 f(\tilde{r}) \chi(\tilde{r})\tilde{r}^3}{\Delta(\tilde{r})} \Big(\frac{4a^2M-\tilde{r}(\tilde{r}-3M)^2}{a^2(\tilde{r}+3M)(\tilde{r}-M)} \Big) \,, \\
    \zeta(\tilde{r})&= \frac{(3 M+\tilde{r}) \Delta (\tilde{r})}{2  \chi(\tilde{r})  \sqrt{a^2 (M+\tilde{r})^2-6 M^2 \tilde{r}^2+2 \tilde{r}^4}}~.
\end{align}
Substituting Eqs.~\eqref{radial deviation},\eqref{eta deviation},\eqref{lambda deviation}, into the radial and angular potentials \eqref{radial pot},\eqref{theta pot}, allows us to compute the symplectic 2-form on the near-critical region of phase space, where the first two terms are given by the critical symplectic 2-form in Eq.~\eqref{eq:critical symplectic form},
\begin{align} \label{eq:near-critical symplectic form}
    \omega &=d\theta \wedge \Big( \frac{\partial p_{\theta}}{\partial \tilde{r}} d\tilde{r}+\frac{\partial p_{\theta}}{\partial E} dE \Big)\\& \nonumber
    +d\phi \wedge \Big(\frac{\partial p_{\phi}}{\partial \tilde{r}} d\tilde{r}+\frac{\partial p_{\phi}}{\partial E} dE\Big)
    \\& \nonumber
    - \frac{\partial p_{r}}{\partial [\delta \rho]} d[\delta \rho] \wedge \Big(d\tilde{r}+d[\delta r]\Big) \\&+(\frac{\partial p_r}{\partial [\delta r]}-\frac{\partial p_r}{\partial \tilde{r}})d\tilde{r} \wedge d[\delta r]-\frac{\partial p_r}{\partial E} dE \wedge (d\tilde{r}+d[\delta r]) \,. \nonumber
\end{align}
Taking $|\delta \rho| \sim \delta r^2 \ll 1$ and working to leading order,
the relevant\footnote{The coefficients $\frac{\partial p_r}{\partial [\delta r]},\frac{\partial p_r}{\partial \tilde{r}}, \frac{\partial p_r}{\partial E}$ are given in Appendix \ref{app:coefficients}; however, they do not contribute to the final expression of the volume form.} coefficients are given in equations \eqref{eq:coefficients ptheta}-\eqref{eq:coefficients pphi} and $ \frac{\partial p_{r}}{\partial \delta \rho}$ is given by
\begin{align}
    \frac{\partial p_{r}}{\partial [\delta \rho]} \approx -\frac{\pm_r E \, \zeta(\tilde{r}) \tilde{r} \sqrt{\chi (\tilde{r})}}{\Delta (\tilde{r}) \sqrt{\delta r^2-\delta \rho \, \zeta(\tilde{r})}}  \,.
\end{align}
The phase-space volume form in the near-photon shell region is therefore given by
\begin{align} \label{eq:dvNPS}
   &dv_{\text{NPS}} = \frac{1}{3!}(\omega \wedge \omega \wedge \omega)\nonumber\\&= \Omega_{\text{NPS}} ( d[\delta \rho] \wedge d
  [\delta r] \wedge d\theta 
  \wedge  d\phi \wedge dE \wedge d\tilde{r}) ~,
\end{align}
where 
\begin{align}
\Omega_{\text{NPS}} &\approx \frac{\pm_r E \, \zeta(\tilde{r}) \tilde{r} \sqrt{\chi (\tilde{r})}}{\Delta (\tilde{r}) \sqrt{\delta r^2-\delta \rho \, \zeta(\tilde{r})}} \Omega_{\text{PS}}~,
\end{align}
and $\Omega_{\text{PS}}$ is defined in equation \eqref{eq:density of states}.
The ordering of exterior products in \eqref{eq:dvNPS} defines positive orientation on the $(\pm_r,\pm_\theta)=(1,1)$ and $(\pm_r,\pm_\theta)=(-1,-1)$ sheets, whereas positive orientation on the $(\pm_r,\pm_\theta)=(-1,1)$ and $(\pm_r,\pm_\theta)=(1,-1)$ sheets is defined by its odd permutation; see discussion in Sec.~\ref{The critical phase space volume}.
The total near-critical phase-space volume can be obtained by performing the $\tilde{r},\theta,\phi,E,\delta \rho, \delta r$ integrations. The range of $\tilde{r},\theta,\phi,E$ are the same as in the case of the critical phase-space submanifold. It is necessary, however, to determine the integration limits for the new coordinates $\delta \rho, \delta r$. 
\subsection{Boundaries of the near-critical phase space}
\noindent
In this section, we define the boundaries of the near-critical phase space, which we will use later as the integration limits of $\delta r$ and $\delta \rho$ in volume computations. To this end, we employ some of the results of Ref.~\cite{gralla2020lensing} concerning the behavior of near-critical light rays.
Define the radial path integral for geodesic motion from point $a$ to point $b$ by
\begin{align}\label{radial path int}
    I^{ab}_r = \fint_{r_a}^{r_b} \frac{dr}{\pm_r \sqrt{\mathcal{R}(r)}}~,
\end{align}
where the slash sign in the integral $\fint$ indicates that the integral is evaluated along the photon trajectory, and in particular the sign $\pm_r$ flips at radial turning points, where the potential $\mathcal{R}(r)$ vanishes.
Expanding the radial potential for near-critical light rays \eqref{eq: delta rho def}, close to the photon shell radius \eqref{radial deviation}, and
keeping $\delta r^2 \sim \delta\rho$ \cite{gralla2019black,johnson2020universal}, yields
\begin{align}
\mathcal{R}(\delta r) =4  \chi(\tilde{r}) \tilde{r}^2 \left(\delta r^2-\delta r^2_{0} \right)~,
\end{align}
where $\delta r^2_{0} =\zeta(\tilde{r}) \delta \rho$. If $\delta r^2_{0}>0$ (equivalently, $\delta \rho>0$), the potential vanishes when $\delta r = \pm \delta r_{0} \equiv \pm \sqrt{\delta r^2_{0}}$: these are the radial turning points.
For photons that reach infinity, only the positive root is relevant. 
In contrast, if $\delta r^2_{0}<0$ (equivalently, $\delta \rho<0$), there are no radial turning points.

Since near-critical photons can propagate all the way to infinity or the horizon, there is an arbitrariness in the definition of the radial boundary of the near-critical phase space. Here we will choose to define the radial boundary so that the points included in the near-critical phase space are no more than one half-orbit away--in Hamiltonian evolution either forward or back in time--from the point where $|\delta r|$ is minimal, or in other words either $\delta r=\delta r_0$ when there is a radial turning point, or $\delta r=0$ otherwise.
Using the definitions of \cite{gralla2020lensing}, and the equations of motion, the fractional number of orbits $n_{ab}$ executed in the motion from $a$ to $b$ is related to the radial path integral via
\begin{align}
 I^{ab}_r \equiv G^{ab}_{\theta} \,,
\end{align}
where the geodesic polar path integral is given by
\begin{align}\label{eq: G_theta}
    G^{ab}_{\theta} \equiv \int_{\theta_a}^{\theta_b} \frac{d\theta}{\pm_{\theta} \sqrt{\Theta(\theta)}}~,
\end{align}
and the fractional number of orbits $n_{ab}$ is defined by
\begin{align}\label{eq: definition of n}
    n_{ab} = \frac{G^{ab}_{\theta}}{ G^{\text{1 orbit}}_{\theta}} \,,
\end{align}
where one normalizes its value over one half orbit
\begin{align}\label{eq: G_theta one half orbit}
     G^{\text{1 orbit}}_{\theta} = 2\int_{\theta_-}^{\theta_+} \frac{d\theta}{\sqrt{\Theta(\theta)}} =\frac{4 K[\frac{u_+}{u_-}]}{a \sqrt{-u_-}}~,
\end{align}
hence we obtain
\begin{align}\label{radial deviation relation to n}
    I^{ab}_r &= \frac{4 K[\frac{u_+}{u_-}] }{a \sqrt{-u_-}} \,n_{ab} \,,
\end{align}
where $u_{\pm}$ are given in equation (\ref{upm}). 
Since we defined the radial boundary of the near-critical phase space to lie one half-orbit away from the points of minimal $|\delta r|$ of a near-critical geodesic, it is clear that for large enough $n$, $\delta r \ll M$ in the entire $n^{\text{th}}$ slice of the near-critical phase space. In this case, the radial path integral simplifies and gives
\begin{align}\label{I Outside}
    I^{ab}_r =
    \frac{1}{2 \tilde{r} \sqrt{\chi(\tilde{r})}}\left. \arccosh\left(\frac{\delta r}{\delta r_0}\right) \right|^{\delta r_b}_{\delta r_a} \, ,
\end{align}
for $\delta r^2_{0}>0$, and
\begin{align}\label{I Inside}
    I^{ab}_r =\frac{1}{2 \tilde{r} \sqrt{\chi(\tilde{r})}}\left.\arcsinh\left(\frac{\delta r}{ |\delta r_0| }\right) 
\right|^{\delta r_b}_{\delta r_a}~,
\end{align}
for $\delta r^2_{0}<0$.

In the case $\delta r^2_{0}>0$, the lower radial boundary is simply the turning point, $\delta r_-=\delta r_{0}=\sqrt{\zeta(\tilde{r}) \delta \rho}$. The upper radial boundary is given by setting $n_{ab}=\pm 1/2$, which yields $\delta r_+ = \sqrt{\zeta(\tilde{r}) \delta \rho}\cosh{( \gamma)}$.
In the case $\delta r^2_{0}<0$, there is no turning point; taking $\delta r_b=0$ and setting $n_{ab}= 1/2$ gives the lower boundary $\delta r_- = -\sqrt{-\zeta(\tilde{r}) \delta \rho}\sinh{( \gamma)}$, and taking $\delta r_a=0$ and setting $n_{ab}= 1/2$ gives the upper boundary $\delta r_+ = \sqrt{-\zeta(\tilde{r}) \delta \rho}\sinh{( \gamma)}$.
Next, we will determine the limits for $\delta \rho$. 
Here, there is a natural choice for relating the deviation in the impact parameter $\delta \rho$ with the half-orbit number $n$; namely, through the complete integral $I_r$ along the entire trajectory. For fixed $\tilde{r}$, a choice of $\delta \rho$ will determine the total fractional number of half-orbits executed by the geodesic. Inverting this relation allows us to determine the outer/inner boundaries (corresponding to larger/smaller $\rho$, respectively), on the $\delta \rho$ axis, of the intervals in which the integer part of the total half-orbit number of a geodesic assumes a particular value $n$.
Explicitly, rewriting the results of \cite{gralla2020lensing} for $\delta \rho>0$, 
\begin{align} 
    I^{\mathrm{tot}}_r &= -\frac{1}{2 \tilde{r} \sqrt{\chi(\tilde{r})}} \log{\Big[\Big(\frac{1+\sqrt{\chi(\tilde{r})}}{8\chi(\tilde{r})}\Big)^2
    \frac{\zeta(\tilde{r}) \delta \rho}{\tilde{r}^2}\Big]}  \,,
\end{align}
while for $\delta \rho<0$,
\begin{align}
    I^{\mathrm{tot}}_r &= -\frac{1}{2 \tilde{r} \sqrt{\chi(\tilde{r})}} \Big(\log{\Big[\frac{\sqrt{1-\chi(\tilde{r})}(1+\sqrt{\chi(\tilde{r})}}{(8\chi(\tilde{r}))^2}} \Big]\\& \nonumber +\log{\Big[ \sqrt{\frac{1+Q_2(z_{+},0)}{1-Q_2(z_{+},0)}}
    \frac{\zeta(\tilde{r}) (-\delta \rho)}{\tilde{r}^2}\Big]} \Big) \,,
\end{align}
where
\begin{align}
    Q_2(z_1, z_2) &= \frac{2 \sqrt{Q(z_1)} \sqrt{Q(z_2)}}{Q(z_1)+Q(z_2)-\frac{(z_1 - z_2)^2}{4 \tilde{r}^2\chi}} \,,\\
    Q(z) &= 1+ \frac{z}{\chi \tilde{r}}+\frac{z^2}{4 \chi \tilde{r}^2} \,,\\
    z_+ &= M + \sqrt{M^2-a^2}-\tilde{r} \,.
\end{align}
Inverting these relations, we find that the $n^\mathrm{th}$ slice of the near-critical phase space is composed of two disconnected parts; for $\delta \rho>0$, the outer/inner boundaries are, respectively,
\begin{align}
    \delta \rho_\pm = \frac{\tilde{r}^2}{\zeta(\tilde{r})} C_{[\delta\rho > 0]} e^{-2 \gamma k_\pm} \, ,
\end{align}
with $k_+=n$, $k_-=n+1$, and for $\delta \rho<0$, the outer/inner boundaries are, respectively,
\begin{align}
    \delta \rho_\pm = - \frac{\tilde{r}^2}{\zeta(\tilde{r})} C_{[\delta\rho < 0]} e^{-2 \gamma k_\mp} \, ,
\end{align}
where
\begin{align}
     C_{[\delta\rho > 0]} &=\left[\frac{8\chi(\tilde{r})}{1+\sqrt{\chi(\tilde{r})}}\right]^{2} \,, \\
      C_{[\delta\rho < 0]} &=  \frac{(8\chi(\tilde{r}))^2}{\sqrt{1-\chi(\tilde{r})}\left(1+\sqrt{\chi(\tilde{r})}\right)} \sqrt{\frac{1-Q_2(z_{+},0)}{1+Q_2(z_{+},0)}} ~.
\end{align}
Having defined the boundaries of the ($n^\mathrm{th}$ slice of the) near-critical phase space, we can proceed and integrate over all phase-space coordinates except for $d\tilde{r}$ and $dE$. The differential phase-space volume of the $n^\mathrm{th}$ slice is given by\footnote{Including also photons that end up at the BH horizon would have given an additional factor of 2 to the near-critical phase-space volume.}
\begin{align}\label{eq: near critical phase space}
&\frac{dV_{\mathrm{NPS}}}{dE d\tilde{r}}
 = \frac{32 \pi \tilde{r}^4  \gamma(\tilde{r})  E^2(C_{[\delta\rho < 0]}+C_{[\delta\rho > 0]})}{a^2 \sqrt{-u_{-}} (\tilde{r}-M) \Delta (\tilde{r})} \nonumber \\& \nonumber
  \times ( e^{-2 \gamma(\tilde{r})  (n+1)}(e^{2 \gamma(\tilde{r})}-1) \chi (\tilde{r})^{\frac{3}{2}} ) \\& \nonumber \Big[ a^2 u_{-} (\tilde{r}-M) \left(K\Big[\frac{u_{+}}{u_{-}}\Big]-E \Big[\frac{u_{+}}{u_{-}}\Big]\right)\\& +\tilde{r}^2 (3 M+\tilde{r}) K\Big[\frac{u_{+}}{u_{-}}\Big]\Big]~,
\end{align}where $u_{-}<0$ and $u_{\pm}$ are given in equation (\ref{upm}). Therefore, the full near-critical phase-space volume of the $n^\mathrm{th}$ slice is given by
\begin{align}\label{enlarged PS}
      \frac{dV_{\mathrm{NPS}}}{dE} = \int_{\tilde{r}_-}^{\tilde{r}_+}  \frac{dV_{\mathrm{NPS}}}{dE d\tilde{r}} d\tilde{r} \,.
\end{align}
It is instructive to take the Schwarzschild limit of \eqref{enlarged PS}. Similarly to the computation described in Sec.~\ref{sec: schw limit of critical}, one must carefully expand the integrand and integration limits for $a \ll M$, integrate, and only then take the $a/M \to 0$ limit of the result. 
In this limit, all integrations can be performed analytically and one finds that the total phase-space volume is given by 
\begin{align}\label{SW approx PS}
  \frac{dV^{\mathrm{Schw}}_{\mathrm{NPS}}}{dE} &= (11 \sqrt{3}-1)(2^7)(3^8)(M \pi)^3 E^2 \\& \nonumber \times e^{- 2\pi (n+1)}(e^{2 \pi}-1)~.
\end{align}
This result can be checked numerically by taking small values of $a/M$ in \eqref{enlarged PS}. 
It would be interesting to consider in detail also the near-extremal limit of the near-critical phase-space volume. A detailed treatment of this double limit, which appears to be more subtle than the zero-spin limit, is beyond the scope of the present paper.
\subsection{Approximate invariance of the near-critical differential phase space volume under Hamiltonian-preserving changes of time foliation}
\noindent
In this section we will show that the near-critical phase space volume in Eq.~\eqref{eq: near critical phase space} is approximately invariant under changes of time foliations of the form given in Eqs.~\eqref{eq: time foliation change},~\eqref{eq: momentum transformation} up to corrections of order $O(e^{-3\gamma n})$. More precisely, for any smooth choice of time-independent change of time foliation $f$, there exists some $n_0$ such that the correction to the near-critical phase-space volume is small for all $n>n_0$.
As in Sec.~\ref{Invariance}, the spacetime coordinate transformations we consider are of the form
\begin{align}
    \bar{t} &= t - f(r,\theta, \phi) \quad ; \quad x^{\bar{i}} = x^i~,
\end{align}
where $ i \in \{r,\theta,\phi\}, \bar{i} \in \{\bar{r},\bar{\theta},\bar{\phi},\}$ and $f(r,\theta, \phi)$ is a continuously differentiable function of the spatial coordinates. 
We again assume that $f$ is time-independent, in order to avoid the introduction of explicit time dependence to the metric and Hamiltonian. The momenta transformations are
\begin{align}
    p_{\bar{t}}=p_t \quad ; \quad p_{\bar{i}}=p_i+p_t \partial_i f~.
\end{align}
We now expand $f$ about $r=\tilde r$. In the near-critical region, $r=\tilde r+\delta r$ with $\delta r \ll M$, and hence
\begin{align}
    f(\tilde{r}+\delta r,\theta, \phi) = f(\tilde{r},\theta, \phi) +\delta r \left.\frac{\partial f}{\partial r}\right|_{r=\tilde{r}} +O(\delta r^2)~,
\end{align}
Since $\delta r_{\pm} \sim e^{-\gamma(\tilde{r})n/2}$, the contribution from the $O(\delta r)$ term to the radial integral is,
\begin{align}
  &\left.\frac{\partial f}{\partial r}\right|_{r=\tilde{r}} \int_{\delta r_-}^{\delta r_+} (\delta r) d[\delta r] =\frac{1}{2}(\delta r^2_+ - \delta r^2_-)\left.\frac{\partial f}{\partial r}\right|_{r=\tilde{r}}~,
\end{align}
%Since $\delta r_{\pm} \sim e^{-\gamma(\tilde{r})n/2}$, this contribution is exponentially suppressed in $n$,
where,
\begin{align}
\frac{1}{2}
\left(\delta r_+^2-\delta r_-^2\right)
=
O\left(e^{-\gamma(\tilde{r})n}\right)~.
%\xrightarrow{n\to\infty}0~,
\end{align}
In addition, the integration range in $\delta \rho$ is of order $O\left(e^{-2 \gamma n}\right)$, since $\delta \rho_{\pm} \sim e^{-2 \gamma(\tilde{r}) n}$. Thus, the resulting correction to the near-critical phase space volume is of order
\begin{align}
   O\left(e^{-2 \gamma n}\right) O\left(e^{-\gamma n}\right)= O\left(e^{-3 \gamma n}\right)~,
\end{align}
which is subleading compared with the leading near-critical phase space volume, $O\left(e^{-2\gamma n}\right)$.
%provided the relevant derivatives of $f$ remain bounded in a neighborhood of the critical photon shell. Therefore, the $O(\delta r)$ correction remains exponentially suppressed as $n\to\infty$. To the order considered here, we may thus evaluate $f$ and its angular derivatives at $r=\tilde{r}$, neglecting their dependence on $\delta r$.
The canonical symplectic 2-form in the new coordinate system is given by Eq.~\eqref{eq: new symplectic 2-form}.
 The third power of the new symplectic form gives the new volume form,
\begin{align}\label{eq:new volume form near-critical}
    \bar{\omega} \wedge \bar{\omega} \wedge \bar{\omega} = \omega \wedge \omega  \wedge \omega + 3\omega \wedge \omega \wedge df \wedge dE~.
\end{align}
Contracting Eq.~\eqref{eq:new volume form near-critical} with $i_{\xi_{E}},i_{\xi_{\tilde{r}}}$ gives
\begin{align} \label{eq: barred differential NPSV}
&\frac{d\mathrm{\bar{V}}_\mathrm{NPS}}{dE d\tilde{r}}(\bar{E},\bar{\tilde{r}}) = \frac{d V_\mathrm{NPS}}{dE d\tilde{r}}(\bar{E},\bar{\tilde{r}}) \\ 
&-6\int_{x \in \mathcal{M}_{\mathrm{NPS}} , E(x) = \bar{E}, \tilde{r}(x)=\bar{\tilde{r}}} \frac{\partial p_r}{\partial [\delta 
\rho]}\Big(\frac{\partial p_{\theta}}{\partial \tilde{r}} \frac{\partial f}{\partial \phi} -\frac{\partial p_{\phi}}{\partial \tilde{r}} \frac{\partial f}{\partial \theta}  \Big) \\& ~~~~~~~~~~~~~~ \times d[\delta r] \wedge d[\delta \rho] \wedge d\theta \wedge d \phi  \,, 
\end{align}
where $\frac{d V_\mathrm{NPS}}{dE d\tilde{r}}$ is given in Eq.~\eqref{eq: near critical phase space}.
Let's focus on the correction term,
\begin{align}
    \int_{x \in \mathcal{M}_{\mathrm{NPS}} , E(x) = \bar{E}, \tilde{r}(x)=\bar{\tilde{r}}} \delta \Omega,
\end{align}
where
\begin{align}
    \delta \Omega =&\frac{\partial p_r}{\partial [\delta 
\rho]}\Big(\frac{\partial p_{\theta}}{\partial \tilde{r}} \frac{\partial f}{\partial \phi} -\frac{\partial p_{\phi}}{\partial \tilde{r}} \frac{\partial f}{\partial \theta}  \Big) \\& ~~~~~~~~~~~~~~  \times d[\delta r] \wedge d[\delta \rho] \wedge d\theta \wedge d \phi  \, .
\end{align}
On the fixed $(E, \tilde{r})$ submanifold, we have the product decomposition,
\begin{align}
    \mathcal{M}^{(E, \tilde{r})}_{\mathrm{NPS}} \cong \mathcal{M}_{\mathrm{2D}} \times \mathcal{M'}_{\mathrm{2D}}~,
\end{align}
with coordinates $(\delta r, \delta \rho)$ and $(\theta, \phi)$, respectively. Denote the canonical projections onto the two factors by $\pi_1$ and $\pi_2$, so that
\begin{align}
    \pi_1 &: \mathcal{M}_{\mathrm{2D}} \times \mathcal{M'}_{\mathrm{2D}} \to \mathcal{M}_{\mathrm{2D}}~,\\
    \pi_2 &: \mathcal{M}_{\mathrm{2D}} \times \mathcal{M'}_{\mathrm{2D}} \to \mathcal{M'}_{\mathrm{2D}}~.
\end{align}
To leading order in $\delta r$, we can therefore write
\begin{align}
    \delta \Omega = \pi^{*}_1 \alpha_{(\bar{E},\bar{\tilde{r}})} \wedge \pi^{*}_2 \beta_{(\bar{E},\bar{\tilde{r}})}+ O(\delta r)~,
\end{align}
where
\begin{align}
    \alpha_{(\bar{E},\bar{\tilde{r}})} &=\frac{\partial p_r}{\partial [\delta 
\rho]}d[\delta r] \wedge d[\delta \rho] \,, \\
\beta_{(\bar{E},\bar{\tilde{r}})} &= \Big(\frac{\partial p_{\theta}}{\partial \tilde{r}} \frac{\partial f}{\partial \phi} -\frac{\partial p_{\phi}}{\partial \tilde{r}} \frac{\partial f}{\partial \theta}  \Big) d\theta \wedge d \phi  ~.
\end{align}
Now, by integration over the product manifold,
\begin{align}
    \int_{ \mathcal{M}^{(E, \tilde{r})}_{\mathrm{NPS}} } \delta \Omega = \Big(\int_{\mathcal{M}_{\mathrm{2D}}} \alpha_{(\bar{E},\bar{\tilde{r}})} \Big) \Big(\int_{\mathcal{M'}_{\mathrm{2D}}} \beta_{(\bar{E},\bar{\tilde{r}})} \Big) +O\left(e^{-3\gamma n}\right)~.
\end{align}
Here, $\mathcal{M'}_{\mathrm{2D}}$ is precisely the corresponding two-dimensional submanifold of the critical photon shell, and $\beta_{(\bar{E},\bar{\tilde{r}})}$ is exactly the correction term appearing in Eq.~\eqref{eq: barred differential PSV} of Sec.~\ref{Invariance}, where we established the invariance of the critical phase-space volume. Thus, using the result of that section,
\begin{align}
    \int_{\mathcal{M'}_{\mathrm{2D}}} \beta_{(\bar{E},\bar{\tilde{r}})} =0
\end{align}
Thus, the leading contribution vanishes, and we obtain
\begin{align}
\int_{\mathcal{M}^{(E, \tilde{r})}_{\mathrm{NPS}}}
\delta \Omega
=
O\left(e^{-3\gamma n}\right)~.
\end{align}
On the other hand, the near-critical phase-space volume itself is of order
$O(e^{-2\gamma n})$. Therefore, the correction is subleading by a relative factor of
$O(e^{-\gamma n})$, which vanishes as $n\to\infty$. Hence, the near-critical phase-space volume is invariant to leading order in the near-critical expansion, with exponentially suppressed subleading corrections. For the near-critical phase space volume differential in energy only, the argument proceeds similarly. We have verified that the same argument applies in this case, but do not present the details here.
%Hence, the leading correction to the near-critical phase-space volume vanishes, while the remaining correction is exponentially suppressed as $n\to\infty$. Therefore, the near-critical phase-space volume is invariant up to corrections of order $O(e^{-3\gamma n})$.}
\section{Discussion} \label{sec:conclusions}
\noindent The BH photon shell is a codimension-two invariant submanifold of the null-geodesic phase space, consisting of all phase-space points corresponding to critical, bound photon trajectories. Here, we studied the symplectic geometry it naturally inherits from the full null-geodesic phase space. In particular, we constructed the volume form and used it to compute the shell's radially differential phase-space volume (equivalently, the radial density of states) and total volume at fixed energy. We have shown the latter two quantities are invariant under Hamiltonian-preserving changes of time foliation.
We investigated the $a \to 0$ and $a \to M$ limits of our results by analytically expanding around these points. Taking the Schwarzschild limit requires a careful expansion in $a/M\ll1$ since in this limit the radial coordinate degenerates. In the high-spin limit, as is often done in spacetime analyses, taking the near-extremality parameter $\kappa \to 0$ can be combined with a near-horizon scaling limit of the radial coordinate. We find that only two such scaling regions, or ``NHEK bands'' contribute finite phase-space volume in the limit: $(\tilde{r}-r_+)\sim\kappa^0$ (far region) and $(\tilde{r}-r_+)\sim\kappa^1$ (near-NHEK region), with all intermediate scalings contributing vanishing volume. We interpret this as a bifurcation of the shell in the limit, and find that the near-horizon piece accounts for $\approx3\%$ of the total volume. Then, motivated by potential observational applications, we studied the near-critical phase space. The phase space is foliated by slices of fixed half-orbit number $n$, with $n \to \infty$ toward the photon shell. We defined the boundaries of slices and computed their radial volume density and total volume at fixed energy.

Our work defines and computes novel coordinate-invariant geometric properties of the Kerr solution which are of potential physical relevance.
More concretely, several potential applications come to mind. 
Our results could provide a starting point for a phase space-based statistical approach for interpretation of time-averaged characteristics of photon rings (such as the brightness profile, see \cite{Farah2025} for a recent discussion) in future BH images, e.g., \cite{Doeleman2023,ayzenberg2023fundamental,BHEX2024,BHEXphotonring2024}.
Such an approach would be most effective in cases (if they exist) where the time-averaged distribution of photons near the shell can be simply modeled. Further study is required to develop this idea.
For example, a statistical phase space-based approach has been successfully applied to the gravitational three-body problem \cite{monaghan1976statistical, stone2019statistical,ginat2021analytical,  kol2021flux}. In that case, dynamics is (transiently) chaotic, so the time-averaged distribution on phase space is expected to be uniform (in the relevant regimes).
In our system, Kerr null geodesics constitute an integrable system, of course; however, it seems plausible that a stochastic source will create a fairly smooth time-averaged phase-space distribution of photons near the shell; homogeneity in some of the phase-space coordinates is in fact implied by symmetry. It would be interesting to study this distribution in simulations or otherwise.

Our work could also be connected to the theory of BH QNMs. Specifically, in Ref.~\cite{Yang2013} it was found that the QNM spectrum of Kerr bifurcates near extremality into two families: one that is weakly damped, i.e., its imaginary part has a small magnitude; and another with an order-one imaginary part. The first family is associated with modes that are supported predominantly in the (near-)NHEK region, and the second with modes which are mainly supported in the far region. It would be interesting to examine whether the phase-space bifurcation we uncovered could be used to explain the QNM bifurcation, at least in the eikonal (in addition to the extremal) limit, where there exists an explicit correspondence between QNM wavefunctions and critical null geodesics \cite{Yang2012}.
Finally, it would be interesting to connect the present treatment of the near-critical region of phase space to that carried out in Ref.~\cite{hadar2022holography}. In particular, it would be instructive to express the phase-space conformal symmetries uncovered therein in terms of the coordinates used here in Sec.~\ref{extended phase space volume} and work out their action on near-critical phase-space slices.

\acknowledgements
We are grateful to Barbora Bezd\v{e}kov\'{a}, Barak Kol, Alexandru Lupsasca, Jake Solomon, and Andrew Strominger for helpful discussions. J.S. is grateful to Beckham Myers for his valuable guidance and helpful discussions on the mathematical aspects of this work. This work was supported in part by the Israel Science Foundation (grant No. 2047/23).

\appendix
\section{Analytical argument for the bifurcation of the extreme Kerr photon shell} \label{app: analytical bifurcation}
\noindent
In Sec.~\ref{sec:near-extremal} we found that, in the extremal limit, the only photon shell radii that contribute a finite amount to the phase-space volume are those that scale as $\tilde{r}-r_h\sim\kappa^1$ and $\kappa^0$ .
In this appendix, we provide the details of the analytical argument showing that the contributions of intermediate NHEK bands with $0<p<1$ are vanishingly small in the limit. For this, we will generalize Eq. \eqref{eq:rt near NHEK transformation} and probe the near-horizon region of (near-)extreme Kerr by
\begin{align}\label{eq:NHEK trans r}
    \tilde{r} & \to M(1+\kappa^p \tilde{R}) \,.  
\end{align}
If we have $\tilde{r}_1-r_h \sim \kappa^p$ and $\tilde{r}_2-r_h \sim \kappa^q$, with $r_h = M(1+\kappa)$ and $0<q<p<1$, we want to show that
\begin{align} \label{eq:volume of bands}
    \int_{\tilde{r}_1}^{\tilde{r}_2} \frac{d\V}{dE d\tilde{r}} d\tilde{r} = \int_{\tilde{R}_1}^{\tilde{R}_2} \frac{d\V}{dE d\tilde{R}}   d\tilde{R} ~~ \xrightarrow[\kappa \to 0] ~~ 0 \,,
\end{align}
where the integration limits scale as
\begin{align}
 \tilde{R}_1 &\sim \kappa^0 \,, \\
  \tilde{R}_2 & \sim  \kappa^{q-p}~.
\end{align} 
To show this, we will 
bound the integral in Eq.~\eqref{eq:volume of bands}
from above. 
This can be done by using the triangle inequality for integrals:
\begin{align}
  \left|\int_{\tilde{R}_1}^{\tilde{R}_2}\frac{d\V}{dE d\tilde{R}} d\tilde{R}\right| \leq \int_{\tilde{R}_1}^{\tilde{R}_2} \left|\frac{d\V}{dE d\tilde{R}}\right| d\tilde{R}~,
\end{align}
where up to leading order in $\kappa$
\begin{align}
    \left|\frac{d\V}{dE d\tilde{R}}\right| &\leq  \mathrm{const.} \left( \kappa^p + \frac{\kappa^{2-2p}}{\tilde{R}^{3}} \right)\,,
\end{align}
where we used the triangle inequality. 
Integrating over $\tilde{R}$ gives
\begin{align}
\int_{\tilde{R}_1}^{\tilde{R}_2}\left|\frac{d\V}{dE d\tilde{R}} \right| d\tilde{R} \leq \mathrm{const.} \left(\kappa^q + \kappa^{2-2q}  \right) ~~ \xrightarrow[\kappa \to 0] ~~ 0 \,.
\end{align}
Therefore, in the extremal limit the contributions of the NHEK bands with $0<p<1$ are vanishingly small.
\section{Coefficients in Eq.~\ref{eq:near-critical symplectic form}}\label{app:coefficients}
The coefficients $\frac{\partial p_r}{\partial E}, \frac{\partial p_r}{\partial [\delta r]}, \frac{\partial p_r}{\partial \tilde{r}}$, appearing in the expression for the symplectic form on the near-critical region of the phase space, in Eq.~\eqref{eq:near-critical symplectic form}, are given by:

\begin{align}
    \frac{\partial p_r}{\partial E}&=\frac{\pm_r 2 \tilde{r} \sqrt{\delta r^2 \chi (\tilde{r})-\delta \rho \, \zeta(\tilde{r})}}{\Delta (\tilde{r})} \,, \\
    \frac{\partial p_r}{\partial [\delta r]}&= \frac{\pm_r 2 \delta r E \tilde{r} \sqrt{\chi (\tilde{r})}}{\Delta (\tilde{r}) \sqrt{\delta r^2-\delta \rho  \, \zeta(\tilde{r})}} \,,
\end{align}

\begin{widetext}
  \begin{align}
    \frac{\partial p_r}{\partial \tilde{r}}&=  \frac{\pm_r E}{\tilde{r} (\tilde{r}-M)^3 (3 M+\tilde{r})^3 \Delta (\tilde{r})^4 \sqrt{\chi (\tilde{r})} \sqrt{\delta r^2-\delta \rho \, \zeta(\tilde{r})}} \Big[ \delta r^2 \Delta (\tilde{r})^2 (3 M+\tilde{r})^3 (\tilde{r}^2 (3 M-2 \tilde{r})-a^2 M) \\& \nonumber (\tilde{r} (4 M^2-3 M \tilde{r}+\tilde{r}^2)-a^2 (M+\tilde{r}))+8  \tilde{r} \zeta(\tilde{r})^3 \delta \rho (M-\tilde{r})^3 \chi (\tilde{r})^3 (a^4 (M+\tilde{r}) (3 M^2+3 M \tilde{r}+\tilde{r}^2)\\& \nonumber +a^2 \tilde{r} (-3 M^4-14 M^3 \tilde{r}-10 M^2 \tilde{r}^2-2 M \tilde{r}^3+\tilde{r}^4)-\tilde{r}^4 (-15 M^3-6 M^2 \tilde{r}+6 M \tilde{r}^2+\tilde{r}^3))\Big] \,.
\end{align}  
\end{widetext}
\bibliographystyle{unsrt}
\bibliography{reference}
\nocite{*}
\end{document}